\documentclass{article}
\usepackage{authblk}
\usepackage[square,numbers]{natbib}
\usepackage[a4paper, total={15cm,20cm}]{geometry}
\usepackage{hyperref}
\usepackage{lineno}
\usepackage{acro}
\usepackage{booktabs}
\usepackage{tabularx}
\newcolumntype{h}{>{\hsize=.5\hsize}X}
\newcolumntype{f}{>{\hsize=.25\hsize}X}
\newcolumntype{t}{>{\hsize=.33\hsize}X}
\newcolumntype{z}{>{\hsize=.66\hsize}X}
\usepackage{caption}
\usepackage{subcaption}
\usepackage{interval}
\usepackage{amsmath}
\usepackage{framed,multirow}
\usepackage{amssymb}
\usepackage{gensymb}
\usepackage{latexsym}
\usepackage{url}
\usepackage{xcolor}
\usepackage{cleveref}
\crefname{figure}{Fig.}{Figs.}
\crefname{figure*}{Fig.}{Figs.}
\crefname{table}{Table}{Tables}
\crefname{table*}{Table}{Tables}
\usepackage{graphicx}
\graphicspath{{pics/}{parts/pics/}}

\newcommand{\diag}{\operatorname{diag}}
\newcommand\norm[1]{\left\lVert#1\right\rVert}

\acsetup{
    single = true
}
\DeclareAcronym{ct}{short     = {CT}, long       = {computed tomography}}
\DeclareAcronym{mri}{short    = {MRI}, long      = {magnet resonance imaging}}
\DeclareAcronym{lb}{short     = {LB}, long       = {Laplace-Beltrami}}
\DeclareAcronym{lm}{short     = {LM}, long       = {landmark}}
\DeclareAcronym{cpd}{short    = {CPD}, long      = {coherent point drift}}
\DeclareAcronym{icpd}{short   = {ICPD}, long     = {iterative coherent point drift}}
\DeclareAcronym{lbrp}{short   = {LBRP}, long     = {Laplace-Beltrami regularized projection}}
\DeclareAcronym{osnicp}{short = {\mbox{OS-N-ICP}}, long = {optimal step nonrigid iterative closest points}}
\DeclareAcronym{gpa}{short    = {GPA}, long      = {generalized Procrustes analysis}}
\DeclareAcronym{gpmm}{short   = {GPMM}, long     = {Gaussian process morphable model}}
\DeclareAcronym{psm}{short    = {PSM}, long       = {posterior shape model}}
\DeclareAcronym{pca}{short    = {PCA}, long      = {principal component analysis}}
\DeclareAcronym{wpca}{short   = {WPCA}, long     = {weighted principal component analysis}}
\DeclareAcronym{ppca}{short   = {PPCA}, long     = {probabilistic principal component analysis}}
\DeclareAcronym{svd}{short    = {SVD}, long      = {singular value decomposition}}
\DeclareAcronym{ssm}{short    = {SSM}, long      = {statistical shape model}}
\DeclareAcronym{pdm}{short    = {PDM}, long      = {point distribution model}}
\DeclareAcronym{bfm}{short    = {BFM}, long      = {Basel face model}}
\DeclareAcronym{cvai}{short   = {CVAI}, long     = {cranial vault asymmetry index}}
\DeclareAcronym{cvi}{short    = {CVI}, long      = {cranial vault index}}
\DeclareAcronym{cr}{short     = {CR}, long       = {cephalic ratio}}
\DeclareAcronym{svm}{short    = {SVM}, long      = {support vector machine}}
\DeclareAcronym{lda}{short    = {LDA}, long      = {linear discriminant analysis}}
\DeclareAcronym{nb}{short     = {NB}, long       = {naïve Bayes}}
\DeclareAcronym{knn}{short    = {kNN}, long      = {k-nearest-neighbors}}
\DeclareAcronym{bdt}{short    = {BDT}, long      = {bagged decision tree}}
\DeclareAcronym{arap}{short   = {ARAP}, long     = {as-rigid-as-possible}}

\DeclareAcronym{2slbrp}{short  = {\mbox{2S-LBRP}}, long  = {two-step Laplace-Beltrami regularized projection}}
\DeclareAcronym{icpdlbrp}{short  = {\mbox{ICPD-LBRP}}, long  = {iterative coherent point drift with Laplace-Beltrami regularized projection}}
\DeclareAcronym{nicpa}{short  = {\mbox{N-ICP-A}}, long  = {nonrigid iterative closest points affine}}
\DeclareAcronym{nicpt}{short  = {\mbox{N-ICP-T}}, long  = {nonrigid iterative closest point translation}}

\title{A statistical shape model for radiation-free assessment and
classification of craniosynostosis}

\author[1]{Matthias Schaufelberger}
\author[3]{Reinald Peter Kühle}
\author[1]{Andreas Wachter}
\author[3]{Frederic Weichel}
\author[2]{Niclas Hagen}
\author[2]{Friedemann Ringwald}
\author[2]{Urs Eisenmann}
\author[3]{Jürgen Hoffmann}
\author[3]{Michael Engel}
\author[3]{Christian Freudlsperger}
\author[1]{Werner Nahm}
\affil[1]{Institute of Biomedical Engineering (IBT), Karlsruhe Institute of Technology (KIT), Kaiserstr. 12, 76137 Karlsruhe, Germany}
\affil[2]{Institute of Medical Informatics, Heidelberg University Hospital, Im Neuenheimer Feld 130.3, Heidelberg, Germany}
\affil[3]{Department of Oral, Dental and Maxillofacial Diseases, Heidelberg University Hospital, Im Neuenheimer Feld 400, Heidelberg, Germany}

\begin{document}

\maketitle

\section*{Abstract}

The assessment of craniofacial deformities requires patient data which is
sparsely available. Statistical shape models provide realistic and synthetic
data enabling comparisons of existing methods on a common dataset.

We build the first publicly available statistical 3D head model of
craniosynostosis patients and the first model focusing on infants younger than
1.5 years. We further present a shape-model-based classification pipeline to
distinguish between three different classes of craniosynostosis and a control
group on photogrammetric surface scans. To the best of our knowledge, our
study uses the largest dataset of craniosynostosis patients in a
classification study for craniosynostosis and statistical shape modeling to
date. 

We demonstrate that our shape model performs similar to other statistical
shape models of the human head. Craniosynostosis-specific pathologies are
represented in the first eigenmodes of the model. Regarding the automatic
classification of craniosynostis, our classification approach yields an
accuracy of 97.8\,\%, comparable to other state-of-the-art methods using both
computed tomography scans and stereophotogrammetry. 

Our publicly available, craniosynostosis-specific statistical shape model
enables the assessment of craniosynostosis on realistic and synthetic data.
We further present a state-of-the-art shape-model-based classification
approach for a radiation-free diagnosis of craniosynostosis.

\newpage

\section{Introduction}\label{sec:intro}

\subsection{Craniosynostosis}

Craniosynostosis is characterized by the premature fusion of skull sutures in
infants and results in irregular growth patterns. The reported prevalence is
three to six cases per 10,000 live
births~\cite{boulet2008,french1990,shuper1985}. Craniosynostosis can occur
isolated (affecting one suture) or non-isolated (affecting multiple sutures).
Syndromic conditions such as Crouzon, Muenke, or Pfeiffer syndrome have
genetic reasons and lead to multi-suture synostosis. These syndromes tend do
show phenotypical craniofacial findings. Unlike syndromic cases, the causes of
isolated craniosynostosis are believed to be multifactorial. Hereditary
conditions and genetic mutations have been identified to cause premature
fusion of specific sutures~\cite{coussens2007}. Symptoms of isolated
craniosynostosis are a deformity of the neurocranium and consecutivley
viscerocranium. Craniosynostosis has been linked to elevated intracranial
pressure~\cite{renier1982} which can lead to reduced brain growth and reduced
neuropsychological development~\cite{kapp-simon2007}.  Depending on
the involved suture, isolated craniosynostosis can be distinguished into
sagittal synostosis (scaphocephaly), metopic synostosis (trigonocephaly),
unilateral coronary synostosis (anterior plagiocephaly), lambda synostosis
(posterior plagiocephaly) and bicoronal synostosis (brachycephaly). Although
brachycephaly includes the synostosis of both coronal sutures, the medical
community counts it among isolated synostosis.  Surgical treatment involves
resection of the synostosis as well as remodeling and reshaping of the cranial
vault. The operation aims to prevent abnormal brain growth, thus enabling a
regular development of skull and face~\cite{judy2018,engel2012}.
Complications during surgery are rare~\cite{bannink2010} and in most cases a
normalized head shape is achieved~\cite{fearon2009}. The most important
differential diagnosis for craniosynostosis are head deformities caused by
positioning without suture fusion. These head deformities most commonly
present as a non-synostotic posterior plagiocephaly. Positioning deformities
are often treated with positioning pillows, helmet therapy or changes in
positioning behavior~\cite{freudlsperger2016}. For further reading,
the reader is referred to~\cite{nagaraja2013}.

As determined by Virchow's Law, the premature closure of a suture limits the
expansion of the skull perpendicular to the fused suture and causes
compensatory growth along the suture, resulting in distinct head
shapes~\cite{persing1989}.  During diagnosis, physicians perform
visual examination, palpation, cephalometric measurements, and medical
imaging. \Ac{ct} imaging is the gold standard for diagnostic imaging as well
as surgical planning and is routinely performed in many craniofacial centers
worldwide. However, this exposes infants to ionizing radiation which should be
avoided~\cite{engel2012}. One alternative imaging method is Black
Bone \ac{mri}~\cite{eley2014,saarikko2020}, which has the
notable drawbacks that during image acquisition, infants are required to be
sedated to prevent them from moving. Sonographic examinations and 3D
photogrammetry are radiation-free and broadly available diagnostic options.
Photogrammetric scans provide inexpensive and fast means to objectively
quantify head shape without exposure to radiation nor the need of sedation.
They are often used to monitor the condition before surgery and the head
development after the operation~\cite{mertens2017}.

\subsection{Statistical shape modeling}

Statistical shape modeling describes the approach to capture variations of 
geometric shapes by statistical methods. In the medical field, applications of 
\acp{ssm} include distinguishing pathological from physiological subjects, 
assisting in surgical planning, or shape classification. \Acp{pdm} are the 
most common type of \acp{ssm} and use a point cloud representation.  
\cite{cootes1992} introduced the idea to construct \acp{pdm} using training 
instances. A key requirement of this approach is that point identifiers across 
training instances share the same morphological meaning. Landmarks are such 
points, but are often only sparsely available, so intermediate points must be 
included to obtain a full model. Accurate correspondence establishment is 
crucial and has a high influence on the model performance, but is at the same 
time difficult to evaluate as no ground truth is available. After dense 
correspondence establishment, a statistical analysis is performed by computing 
sample mean and sample covariance matrix of the training data. Usually, this 
is a \ac{pca} or a variation of it which is applied on the covariance matrix, 
arranging the principal components according to their respective variance in 
the training data. Synthetic, valid shapes are constructed by linear 
combinations in the vector space defined by the principal components and 
constrained by their respective eigenvalues.

Shape models of the human face have become popular types of \acp{pdm}. Some of
the most important publicly available shape models of the human face are the
Basel Face Model~\cite{blanz1999,gerig2017} and the Large Scale Facial
Model~\cite{booth2016}. The Liverpool-York-Head-Model is the first publicly
available shape model of the full human head in both shape and
texture~\cite{dai2017a} and includes a child model constructed from subjects
between 2 and 15 years. \cite{egger2020}~gives a overview about face models, current
trends, applications, and challenges.

\subsection{Assessment and classification of head deformities using shape analysis}\label{subsec:assessment}

Researchers studied anatomic differences between physiological subjects and
patients with craniosynostosis~\cite{fabijanska2015} as well as therapy
outcome~\cite{borghi2020} using \ac{ct} scans on both 2D and 3D approaches.
\cite{ruiz-correa2005} discriminated between scaphocephaly and control subjects
using a \ac{svm} on the sampled 2D contour of the skull on the axial plane of
\ac{ct} scans. \cite{marcus2009} performed a statistical analysis on \ac{ct}
scans of physiological subjects before comparing them to subjects of
scaphocephaly. \cite{mendoza2013} developed a shape analysis framework for the
quantification of diagnostic features of \ac{ct} scans. In their follow-up
publication~\cite{mendoza2014}, they proposed a classification framework to
distinguish between types of craniosynostosis using shape descriptors from an
\ac{ssm} and fusion indices of the ossified sutures extracted from \ac{ct}
scans. They also performed a classification using \ac{lda} and an \ac{svm} on
141 cases, the majority being physiological subjects with an overall accuracy
of 95.7\,\% using a leave-one-out classifier.

Recently, the use of radiation-free 3D stereophotographs for head shape
assessment and classification of craniosynostosis gained momentum.
\cite{dai2017} proposed a 2D \ac{ssm} of craniofacial profile and evaluated
contour changes after surgical intervention of 25 craniosynostosis patients.
\cite{rodriguez-florez2017} evaluated surgical spring-assisted therapy by using
the principal components of an \ac{ssm} to assess surgical parameters.
\cite{meulstee2017} constructed a \ac{pdm} by determining the intersections of
cast rays onto photogrammetric scans of 100 patients and found statistically
significant differences between the control group and both scaphocephaly and
trigonocephaly patients. Regarding the classification of craniosynostosis,
\cite{deJong2020} proposed a dense neural network to classify different types
of craniosynostosis using ray casting to compute the distances between a
center point and the surface of 3D stereophotographs of the head. Using
stratified 10-fold cross validation on 196 subjects, they correctly classified
99.5\,\% of their subjects.

\subsection{Scope of this work} 

Works of previous authors using \acp{ssm} from 3D surface scans showed
promising first results to distinguish features of different head
deformities~\cite{meulstee2017,rodriguez-florez2017,heutinck2021}.
However, no craniosynostosis-specific \ac{ssm} has been made publicly
available, so methods are tested on in-house datasets making quantitative
comparisons between methods difficult. Existing publicly
available \acp{ssm} are trained on healthy adults and
children~\cite{dai2017a}, thus assessing head deformities
in terms of healthy, physiological variations of head shapes.

An \ac{ssm} of infants with and without head deformities can be used to create
a dataset which serves as a baseline for comparison studies for pathological
features, for classification studies, and for mesh manipulation
applications such as patient counseling, shape reconstruction or the creation
of synthetic data with specific pathological features. Active shape model
approaches could use an \ac{ssm} of craniosynostosis patients for facilitating
automated segmentation.\\

\noindent This publication makes the following contributions:
\begin{itemize}

    \item We propose the first publicly available \ac{ssm} of craniosynostosis
    patients using 3D surface scans, including pathology-specific submodels,
    texture, and 100 synthetic instances of each class. It is the first
    publicly available model of children younger than 1.5 years and \ac{ssm}
    of craniosynostosis patients including both full head and texture. Our
    model is compatible with the Liverpool-York head model~\cite{dai2020} as
    it makes use of the same point identifiers for correspondence
    establishment. This enables combining texture and shape of both models.

    \item We present an alternative classification approach directly on the
    parameter vector of our \ac{ssm} composed of 3D photogrammetric surface
    scans. We test five different classifiers on our database consisting of
    367 subjects and achieve state-of-the-art results. To the best of our
    knowledge, we conduct the largest classification study of craniosynostosis
    to date.

    \item We demonstrate two applications of our \ac{ssm}: First, with regards
    to patient counseling, we apply attribute regression as proposed
    by~\cite{blanz1999} to remove the scaphocephaly head shape of a patient.
    Second, for pathology specific data augmentation, we use a generalized
    eigenvalue problem to define fixed points on the cranium and maximize
    changes on face and ears as proposed by~\cite{albrecht2008}. To the best
    of our knowledge, neither of these applications have been applied to
    patients using a craniosynostosis shape model before.

\end{itemize}

\section{Materials and methods}~\label{sec:materialsAndMethods}

\begin{figure}[htbp]
\centering
\includegraphics[width=0.65\textwidth]{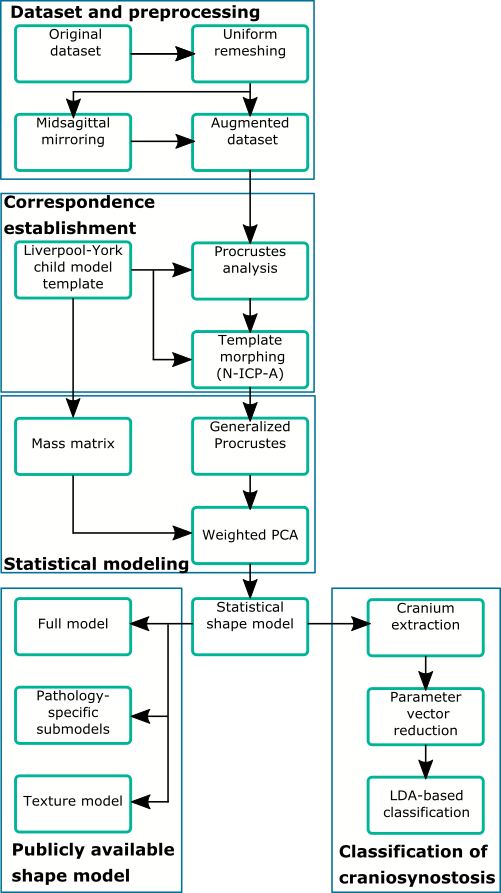}
\caption{Shape model creation and classification pipeline. }\label{fig:schematic}
\end{figure}

\Cref{fig:schematic} gives a full overview of the pipeline from the raw data
to the \ac{ssm} creation and the craniosynostosis classification. We describe
each of the top-level blocks in detail in the following subsections.

\subsection{Dataset and preprocessing}~\label{subsec:dataset} 

At the Department of Oral, and Maxillofacial Surgery of the Heidelberg
University Hospital photogrammetric surface scans are used in the daily
routine to diagnose and document patients with craniofacial diseases. In this
study we were concerned with patients suffering from craniosynostosis. Out of
the scans that were acquired between 2011 and 2020, 367 preoperative 3D
photogrammetric scans were extracted. We used a standardized protocol, which
had been examined and approved by the Ethics Committee Medical Faculty of the
University of Heidelberg (Ethics number S-237/2009). The study was carried out
according to the Declaration of Helsinki and written informed consent was
obtained from parents. The Canfield VECTRA-360-nine-pod system (Canfield
Science, Fairfield, NJ, USA) was used for recording. To avoid artifacts on the
head due to hair, the scanned infants wore tight fitting nylon caps. For each
of the recordings, the dataset provided the 3D vertex coordinates, UV texture
coordinates, and the triangular face indices connecting the vertices to a mesh
surface. Each recording contained additional metadata, which includes the
medical diagnosis of the physician, the patient's age on the day of the
recording, and 10 cranial and facial landmarks manually annotated by a medical
expert. In the appendix, we summarize the aforementioned landmarks
in~\Cref{tab:landmarks}.

\begin{figure}[htb]
\centering
\includegraphics[width=0.9\textwidth]{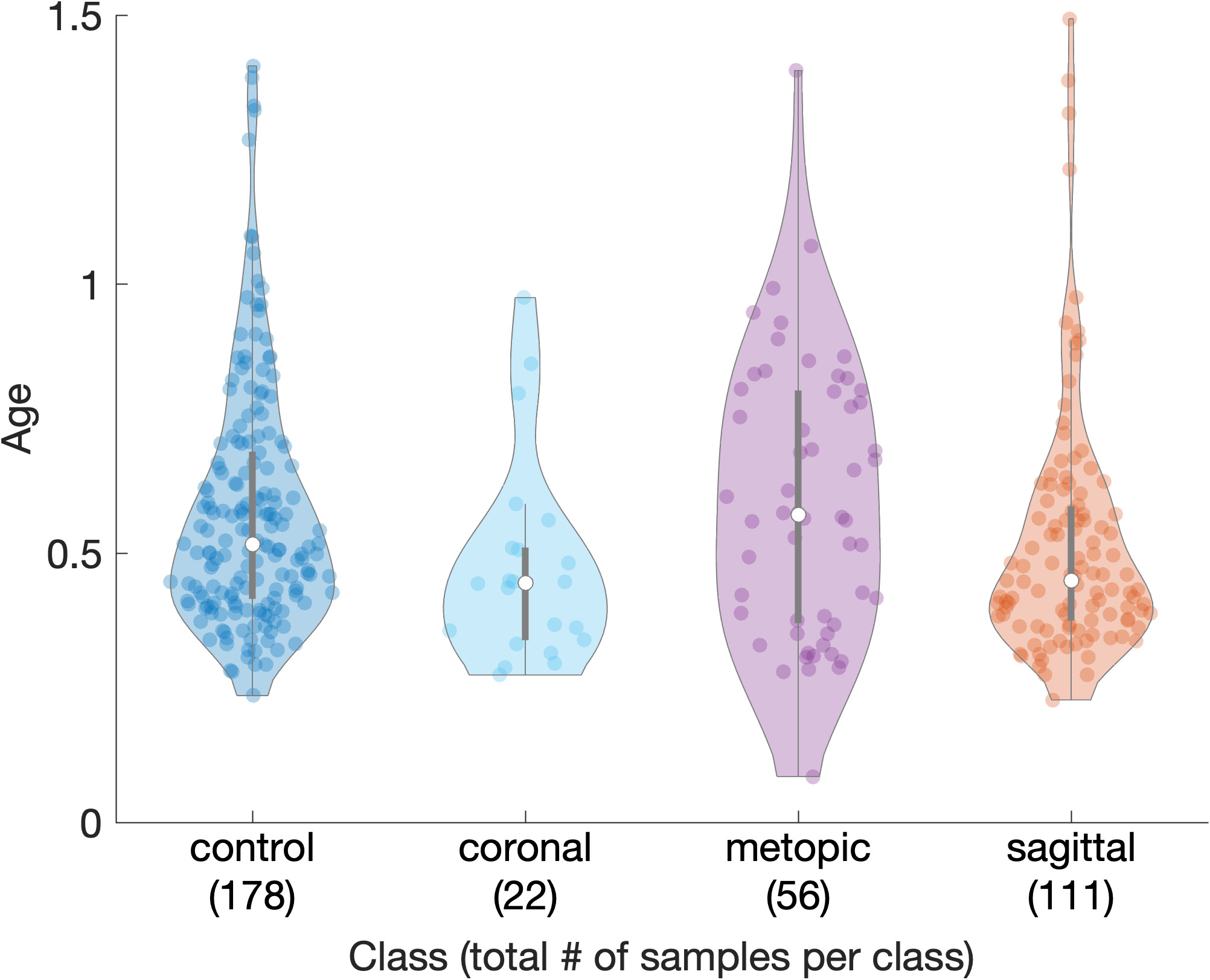}
\caption{Age distribution among classes of the dataset. Parenthesis indicate
number of samples per class.}\label{fig:ageDist}
\end{figure}

We retrieved patient scans classified with three types of craniosynostosis,
namely ``coronal suture fusion'' (brachycephaly and unilateral anterior
plagiocephaly), ``sagittal suture fusion'' (scaphocephaly), ``metopic suture
fusion'' (trigonocephaly), as well as a control group without any suture
fusion. This makes our approach comparable to other classification studies
which distinguished between craniosynostosis and non-craniosynostosis classes,
in particular~\cite{mendoza2014} and~\cite{deJong2020}. Besides healthy
subjects, our control group contained also scans of children with positional
posterior plagiocephaly without suture fusion, who were later treated with
helmet therapy or laying repositioning. All craniosynostosis patients later
underwent surgical remodeling of the cranium. We show violin
plots~\cite{hintze1998} using a publicly available
implementation~\cite{bechtold2021} of the subjects' age distribution
in~\Cref{fig:ageDist}.

During recording, the patients had to be held tight, oftentimes the neck of
the patient was covered by clothes or the hands of the medical staff. For this
reason, some recordings contained isolated parts and other artifacts.
Additionally, close to the ears we often found mesh irregularities such as
large edge lengths. The open source tool \textit{Meshlab}~(ISTI-CNR,
~Pisa,~Italy)~\cite{cignoni2008} was used to remove isolated parts,
duplicated vertices, and to close holes. After artifacts removal we used
isotropic explicit remeshing~\cite{pietroni2010} to avoid large edge lengths
and to obtain a regular, uniform surface scan. 

Depending on the model application, it can be advantageous to use a
non-regular mesh (having different resolutions on different parts of the
mesh). Cranial parts mostly change smoothly in comparison with facial parts
and can thus be expressed with fewer vertices and lower spacial resolution to
reduce computational cost.

We used the mean shape of the Liverpool-York child head model~\cite{dai2017a}
as a template for correspondence establishment. This template has the
advantage of being symmetric and does not have a head deformity avoiding
pathological bias during model creation. To directly include eyes and close
the mouth in the model we added additional vertices into the template which
enables using the model without an additional eye model. The modified template
is compatible with the original model as the order of the vertex identifiers
is the same. The final template has $p = 13151$ vertices and a mean edge
length of 2.91\,mm.

\subsection{Correspondence
establishment}~\label{subsec:correspondenceEstablishment} 

During correspondence establishment we intended to find corresponding points
with the same morphological meaning on the template $\mathbf{X} \in
\mathbb{R}^{p \times 3}$ and the target scans. To increase our sample size and
avoid a symmetry-bias for asymmetric pathological cases for the coronal model,
we mirrored each subject on the midsagittal axis thus increasing the dataset
size from $N = 367$ to $2N = 734$. To prepare template morphing, we initially
aligned the template to each target using the annotated landmarks. We used
Procrustes analysis to obtain a linear transformation consisting of
translation, rotation, and isotropic scaling which transformed the original
template landmarks onto the corresponding target landmarks. By applying the
transformation to the whole set of the template points, we aligned the
template mesh to each target scan, ensuring that the facial and cranial
regions of the template are close to their morphological counterparts on the
target scan. Note that this process only facilitates template morphing and
does neither rescale nor change the target scans.

Numerous registration algorithms for 3D surface registration exist and have
been tested on medical
data~\cite{besl1992,myronenko2010,dai2017a,dai2020,allen2003,booth2016,gerig2017},
many of which can be tailored toward specific applications. We tested and
evaluated four different state-of-the-art methods which have been successfully
used for template morphing of the full head. Two approaches (\Acf{2slbrp},
\acf{icpdlbrp}) use a \acf{lbrp} approach while the two other other methods
(\ac{nicpa}, and \ac{nicpt}) use an \ac{osnicp} approach. After evaluating the
four methods and the resulting models, we used \ac{nicpa} as the final method
for our model. We continue with a description of the \ac{nicpa} method. 

The \ac{osnicp} methods were presented by \cite{amberg2007} who base their work
mostly on~\cite{allen2003}. The core idea is to use an affine transformation
for each point and locally regularize transformations of connected points. A
stiffness term penalizes differences between transformations between adjacent
nodes. A distance term controls how close the template vertices are
transformed to the target points and a landmark term requires that the
landmark points of template and target match each other. All three terms,
stiffness term, distance term, and landmark term, are optimized simultaneously
using an iterative approach starting with a high stiffness. For each
stiffness, a correspondence search is performed and the optimal deformation
with respect to the found correspondences is computed. As soon as the
transformation changes very little, the stiffness parameter is decreased and
repeated for the reduced stiffness until convergence. The mathematical
formulas are presented in~\Cref{ap:osnicpMethods}, but for detailled
explanations, the reader is referred to~\cite{amberg2007}. We used a
exponential descrease of the stiffness parameter and a small landmark term
which was set to zero after 50 iterations. We list all hyperparameters in the
appendix in~\Cref{tab:morphinghyper}. 

A description of the \ac{lbrp}-based methods is presented
in~\Cref{ap:lbrpMethods} and the detailled evaluation and quantitative
comparison of all four methods can be found in~\Cref{ap:morphingComparison}.

\subsection{Statistical modeling} \label{subsec:statisticalModeling} 

To align the morphed templates, we employed rigid \ac{gpa}~\cite{cootes1992}.
\ac{gpa} iteratively calculates the mean shape of the training data, the
deviation of the training data to the calculated mean shape, and aligns the
training data accordingly. The Euclidean distance was used as the Procrustes
distance metric. This removes the non-shape related attributes translation and
rotation from the morphed templates. For this study, we considered scale an
attribute of shape because craniosynostosis-related features may depend on the
patient's age and head size. After reshaping each of our morphed templates
$\mathbf{X}_i \in \mathbb{R}^{p \times 3}$ into columnvectors $\mathbf{x}_i
\in \mathbb{R}^{3p}$, we stacked them horizontally to obtain the observation
matrix $\mathbf{X}_\mathrm{Obs} \in \mathbb{R}^{3p \times 2N}$. We regarded
them as independent observations which served as training data upon which we
built our \ac{ssm}. We computed the mean shape $\bar{\mathbf{x}} \in
\mathbb{R}^{3p}$ and by subtracting the mean shape from the observations
matrix, we obtained the mean-aligned data matrix which we refer to as
zero-mean data matrix:

\begin{equation}
    \mathbf{X}_\mathrm{ObsZM} = \mathbf{X}_\mathrm{Obs} - \bar{\mathbf{x}}
    \label{eq:zeroMean}
\end{equation}

\Ac{pca} is a statistical method to compute the eigenvectors and eigenvalues
of the sample covariance matrix of the zero-mean training data. One drawback
of performing ordinary \ac{pca} is that all points are weighted equally
regardless of their morphological importance. Consequently, areas with higher
vertex density such as the face of our template are therefore over-represented
compared to parts with lower vertex density such as the cranium. As
\cite{dai2019} proposed, we can use \ac{wpca} to counterbalance the influence
of the face by using a mass matrix. 

We defined the mass matrix $\mathbf{M} \in \mathbb{R}^{p \times p}$, composed
of per-vertex weights and per-edge weights in a very similar manner to
barycentric cells. The diagonal elements of $\mathbf{M}$ represent the vertex
weights. Each vertex weight is defined as the sum of the area of the adjacent
faces for which this vertex is the nearest neighbor. Likewise, the
non-diagonal elements represent the edge weights and each edge weight is
defined as the sum of the area of the adjacent faces for which this edge is
the closest edge. To account for the vectorized representation of the
observations, the mass matrix is then stretched by factor $3$ and
nearest-neighbor-interpolated, resulting in $\mathbf{M}_\mathrm{3} \in
\mathbb{R}^{3p \times 3p}$. We computed the weighted Gram matrix
$\mathbf{G}_\mathrm{W} \in \mathbb{R}^{2N \times 2N}$ as
\begin{equation}
\mathbf{G}_\mathrm{W} = \mathbf{X}_\mathrm{ObsZM}^\mathrm{T} \mathbf{M}_\mathrm{3} \mathbf{X}_\mathrm{ObsZM}\label{eq:gram}
\end{equation}
and performed an eigendecomposition of $\mathbf{G}_\mathrm{W}$ using
\begin{equation}
    \label{eq:eigendecomposition}
    \mathbf{G}_\mathrm{W} = \mathbf{U}_\mathrm{G} \mathbf{\Lambda}_\mathrm{G} \mathbf{U}_\mathrm{G}^\mathrm{T}.
\end{equation}
We computed the principal components $\mathbf{V} \in \mathbb{R}^{3p \times 2N}$
of the training data as
\begin{equation}
    \mathbf{V} = \mathbf{X}_\mathrm{ObsZM} \mathbf{U}_\mathrm{G}
    \mathbf{\Lambda}_\mathrm{G}^{-\frac{1}{2}},
    \label{eq:weightedpcaV}
\end{equation}
and the eigenvalues $\mathbf{\Lambda} \in \mathbb{R}^{2N  \times 2N}$ of the
sample covariance matrix of the training data by re-scaling the eigenvalues of
the Gram matrix
\begin{equation}
    \mathbf{\Lambda} = \frac{1}{2N-1} \mathbf{\Lambda}_\mathrm{G}.
    \label{eq:weightedpcaL}
\end{equation}
Each observation could then be defined using the principal components and
eigenvalues using
\begin{equation}
    \mathbf{x} = \bar{\mathbf{x}} + \mathbf{V} \mathbf{\Lambda}^{\frac{1}{2}} \mathbf{\alpha}.
    \label{eq:alpha}
\end{equation}
As the data matrix was centered before, the last eigenvalue will be zero and
can be omitted.

We created one full model, four class-specific submodels, and one cranium-only
model. The full model was created used the full zero-mean observation matrix.
The class-specific submodels used the assigned diagnosis label for each
observation. For the classification approach, we built a cranial model and
extracted the cranial part of the template to remove possible influences of
the face. For each model, \ac{gpa} was performed individually.

\subsection{Classification of craniosynostosis}\label{subsec:classification}

The classifier was trained to distinguish between the labeled classes and thus
between the three different types of craniosynostosis and the control class.
The cranium model served as a basis for the classifiers. We extracted the
coefficient vector $\alpha$ for each observation from the cranial model which
served as an input for the classifiers. Using the coefficient vector as shape
descriptors and as a direct input for an \ac{svm} has been successfully tested
in a different domain~\cite{shen2012}.

We evaluated five different classifiers: \Ac{svm}~\cite{cortes1995},
\ac{lda}~\cite{fisher1936}, \ac{nb}~\cite{zhang2004},
\acp{bdt}~\cite{gordon1984}, and \ac{knn}~\cite{cover1967}. All classifiers
were implemented using the Python module
\textit{scikit-learn}~\cite{pedregosa2011} (version 1.0.2), mostly sticking to
the default settings. \Acp{svm} are binary classifiers that use kernel
functions to map the input parameters into a high-dimensional representation
which can be separated by hyperplanes. We chose a kernel based on radial basis
functions and a multi-class model with $6$ one-versus-one binary classifiers.
For \acp{lda}, we used a multivariate Gaussian distribution for each class
assuming the same covariance matrix for each class. Each prediction is
assigned to the class whose mean is the closest in terms of the Mahalanobis
distance taking into account the prior probability of each class. \Ac{nb}
assumes conditional independence between input variables. Similar to \ac{lda},
we used a Gaussian model to distinguish between classes. \ac{knn}
classification classifies the test sample according to the $k$ closest
neighbors. We selected $k=5$ nearest neighbors in Euclidean space. For
tie-breaking we chose the nearest neighbor among the tied classes. \Acp{bdt}
are white-box classification algorithms using a hierarchical, tree-like
structure. We used the default implmentation for tuning the hyperparameters of
the \acp{bdt}.

We used stratified 10-fold cross validation on the unmirrored samples. For
each split, the test set was only composed of the original, unmirrored samples
and the training set was augmented with the mirrored training samples. This
way, each of the samples from the original set was used once for testing
without the possibility of cross-over.

\Ac{pca} orders the principal components according to their variance, so the
first principal components describe the overall shape while the last
components contain mostly noise. The noise can arise e.g., from incorrect
morphing, limited resolution, or acquisition errors during scanning. We aimed
to reduce the number of principal components based on the assumption that the
parameters responsible for a good classification are concentrated in the first
components. We iterated over the first $100$ principal components and used the
accuracy as a fitness function to select the optimal number of principal
components. Finally, four different metrics evaluated the final classifier:
Beside overall accuracy, we used g-mean, per-class sensitivity, and per-class
specificity.

\section{Results}\label{sec:results}

\subsection{Morphing and model evaluation}\label{subsec:morphingModelEvaluation}

We evaluated each template morphing approach using three metrics: landmark
errors, vertex-to-nearest-neighbor distances, and per-class surface normal
deviations. Landmark errors provide sparse point-to-point errors on known
correspondences. Vertex-to-nearest-neighbor distances evaluate how close the
template has been morphed onto the target without taking into account if the
nearest neighbor is morphologically correct. What we refer to as ``surface
normal deviations'' has been proposed in~\cite{amberg2007}: we removed
translational and rotational components from the morphed templates and
computed surface normal deviations between the morphed templates. This
evaluates how well point identifiers have been morphed onto morphologically
similar regions across all scans. However, our dataset contained different
pathologies, so we also expected shape and surface normal differences among
different pathology classes. Hence, we modified this approach and computed
surface normal deviations on each pathology class separately before
calculating the cumulative mean surface normal deviations. 
\Cref{tab:morphingmeanIcpa} shows morphing errors for the \ac{nicpa} method. 

\begin{table}[htbp]
\caption{Mean error and standard deviation for each morphing method. Boldface
shows smallest error for each metric.}\label{tab:morphingmeanIcpa}
\centering
\small
\begin{tabularx}{0.99\textwidth}{XXX}
\toprule
Mean landmark error (mm) & Mean vertex-to-nearest-neighbor distance (mm) & Mean surface normals deviations (degree) \\
\midrule
$6.533 \pm 1.796$ & $0.007 \pm 0.003$ & $33.488 \pm 1.578$ \\
\bottomrule
\end{tabularx}
\end{table}

For shape model evaluation we used the three metrics compactness,
generalization, and specificity~\cite{styner2003,davies2002}. Compactness
determines the model's ability to capture most of the variance with few
components, generalization the model's ability to fit to unknown observations,
and specificity the model's ability to create synthetic instances similar to
the training data. 

In this section, we only show the results for~\ac{nicpa}, the final morphing
method. All morphing methods are compared in \Cref{ap:morphingComparison}.

\newcommand{\pcscale}{0.14}
\begin{figure}[htbp]
    \begin{center}
        \begin{subfigure}[c]{\textwidth}
        \centering
        \includegraphics[scale=\pcscale]{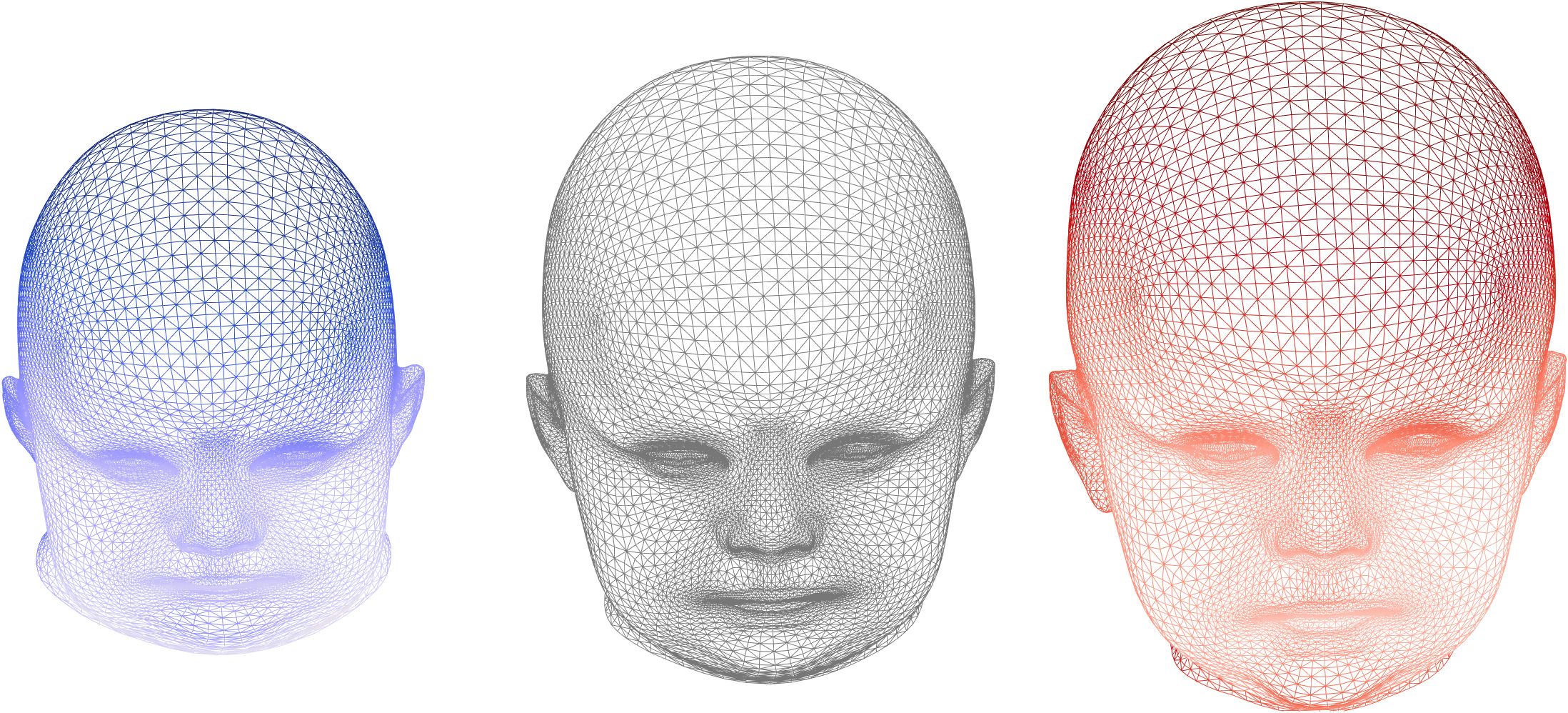}
        \end{subfigure}
        \begin{subfigure}[c]{\textwidth}
        \centering
        \includegraphics[scale=\pcscale]{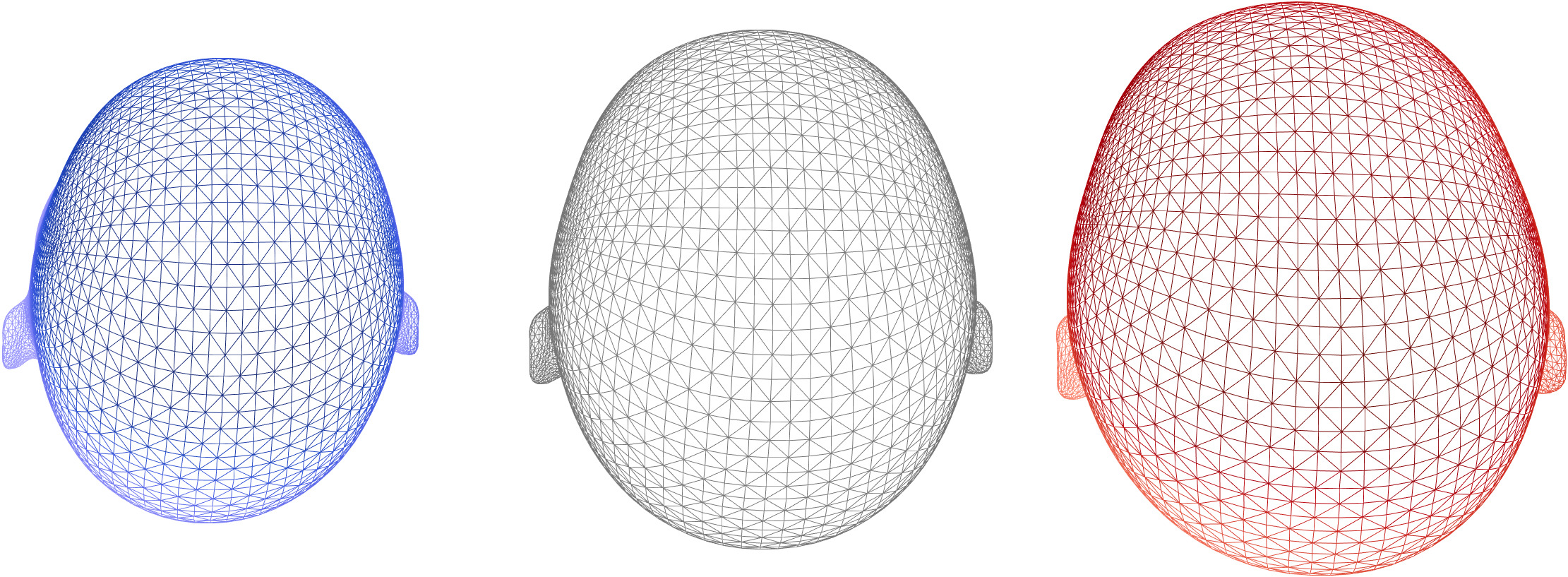}
        \end{subfigure}
        \caption*{\textsf{First mode, front and top view.}}
        \vspace{0.25cm}
        \begin{subfigure}[c]{\textwidth}
        \centering
        \includegraphics[scale=\pcscale]{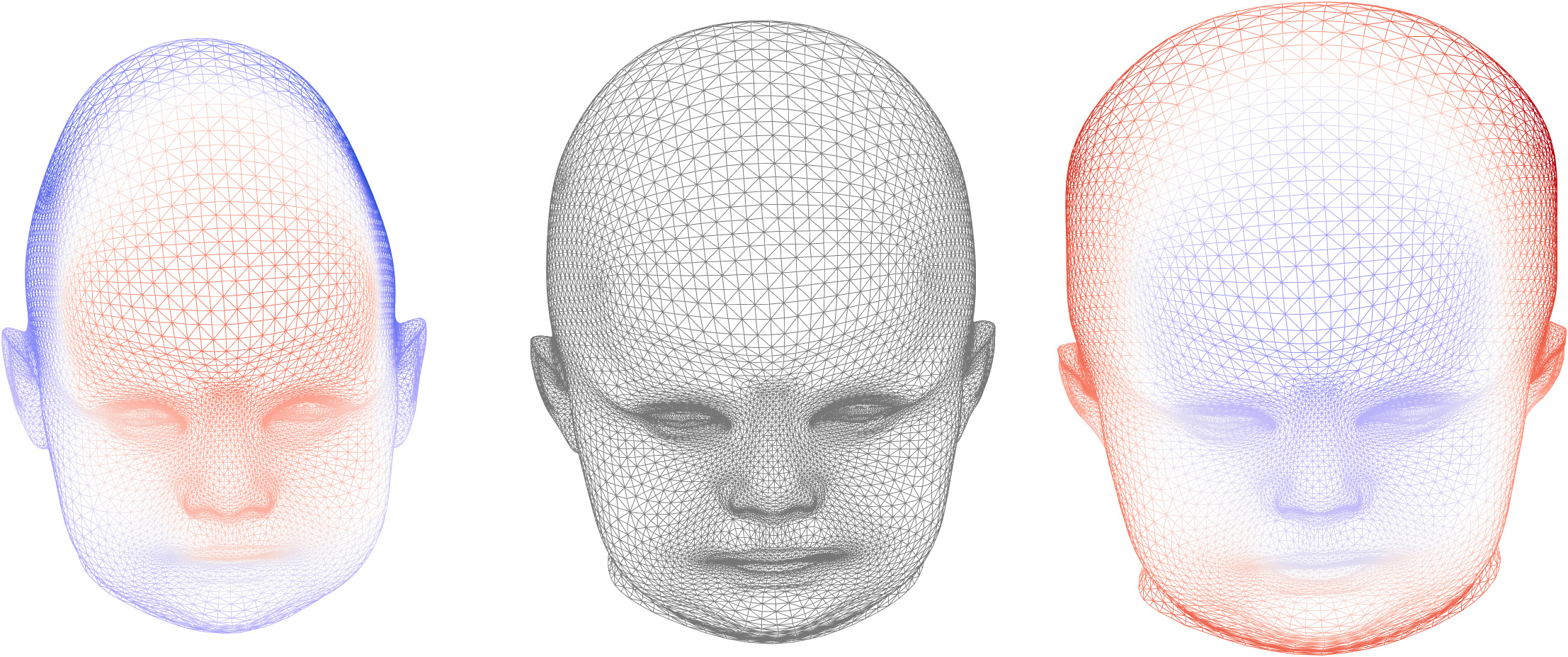}
        \end{subfigure}
        \begin{subfigure}[c]{\textwidth}
        \centering
        \includegraphics[scale=\pcscale]{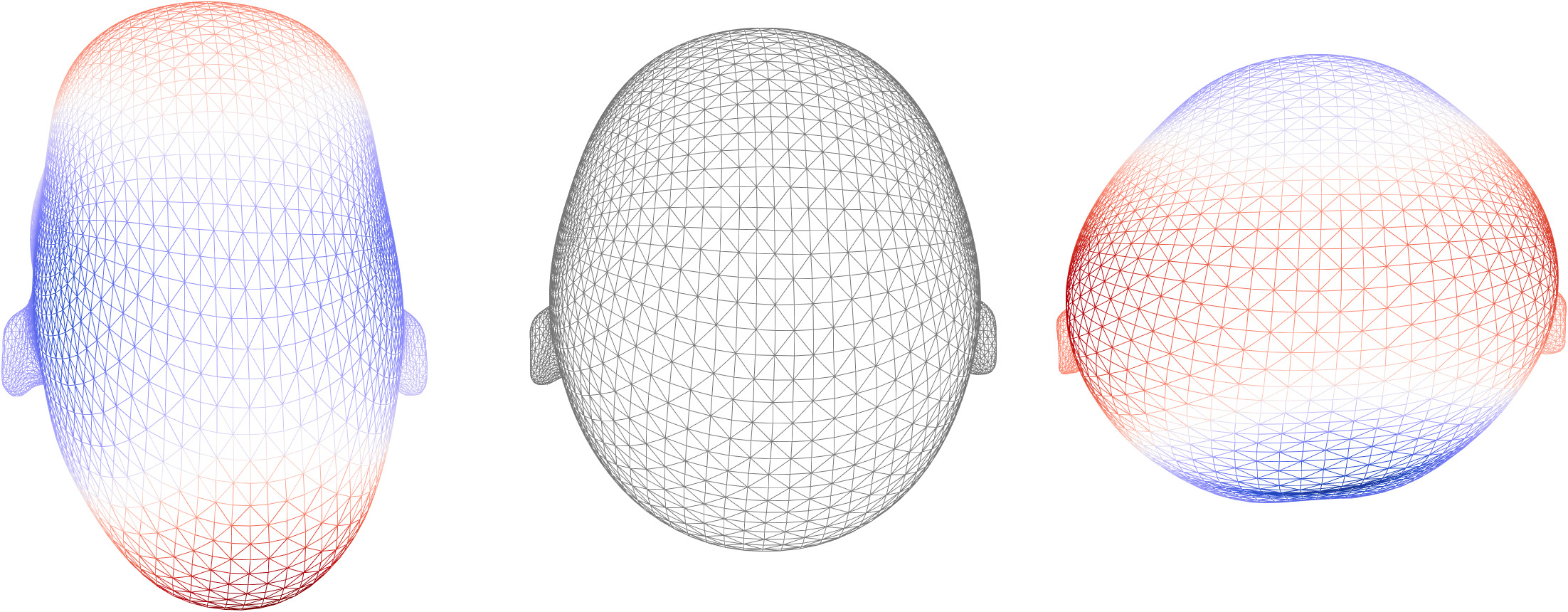}
        \end{subfigure}
        \caption*{\textsf{Second mode, front and top view.}}
        \vspace{0.25cm}
        \begin{subfigure}[c]{\textwidth}
        \centering
        \includegraphics[scale=\pcscale]{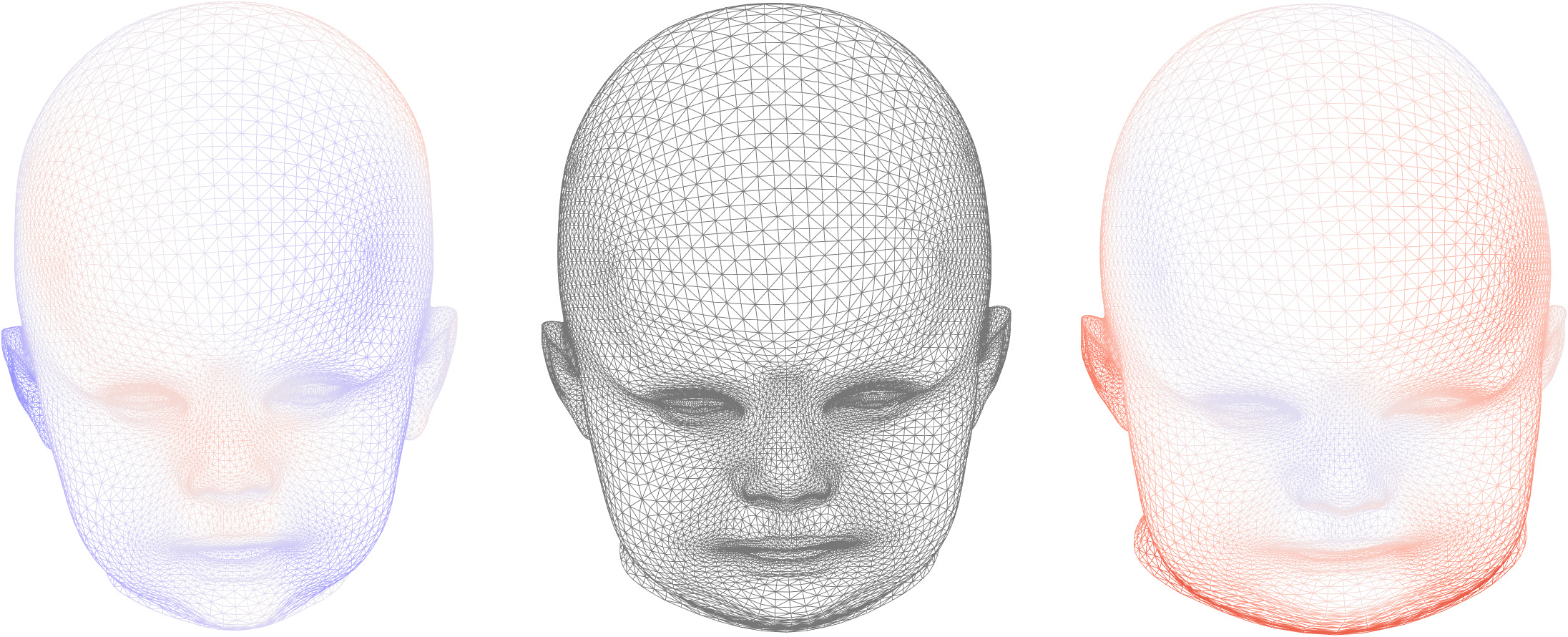}
        \end{subfigure}
        \begin{subfigure}[c]{\textwidth}
        \centering
        \includegraphics[scale=\pcscale]{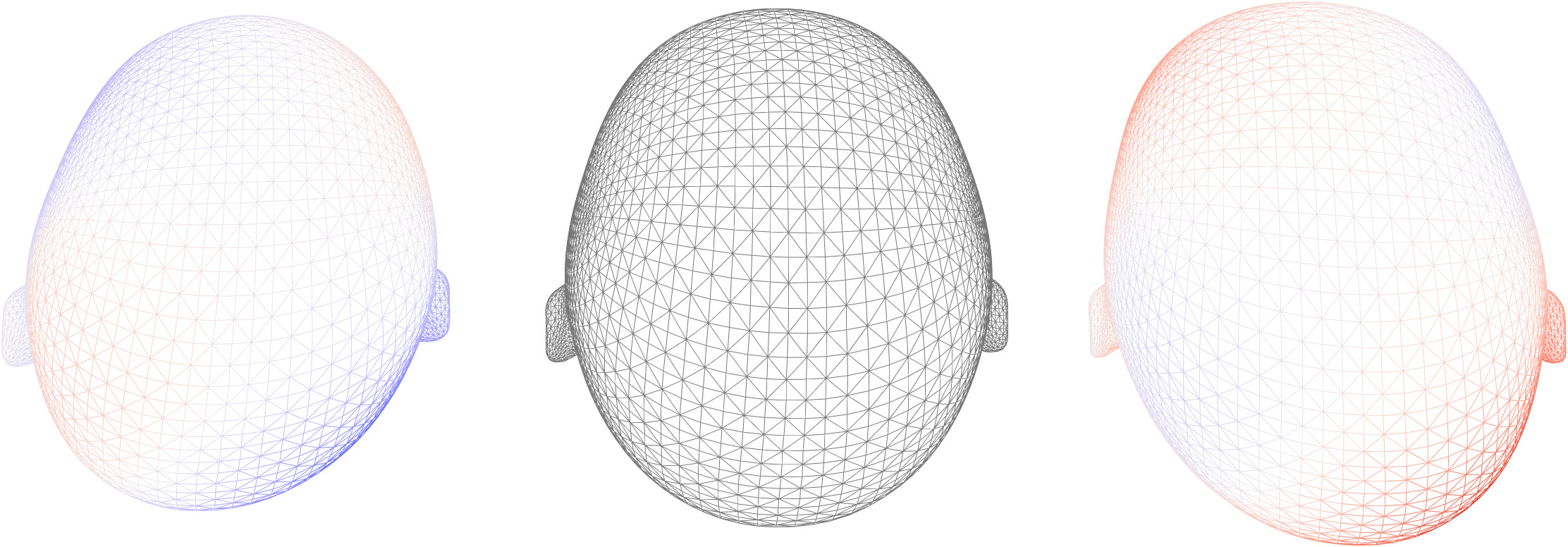}
        \end{subfigure}
        \caption*{\textsf{Third mode, front and top view.}}
        \vspace{0.25cm}
        \begin{subfigure}[c]{\textwidth}
        \centering
        \includegraphics[width=0.4\textwidth]{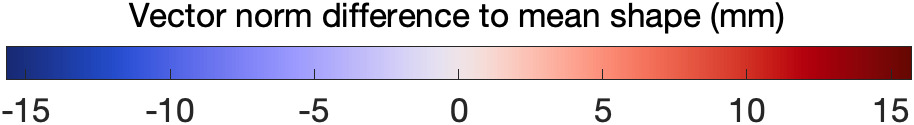}
        \end{subfigure}
        \caption{First three modes of the full model in front and top view. From
        left to right: $-3\sigma$, mean shape, and $+3\sigma$. Colorbar indicates
        vector norm difference between principal component shape and mean shape
        (gray).}
        \label{fig:principalComponents}
    \end{center}
\end{figure}

\newcommand{\compheightIcpa}{0.29}
\begin{figure}[ht]
    \centering
    \includegraphics[height=\compheightIcpa\textwidth]{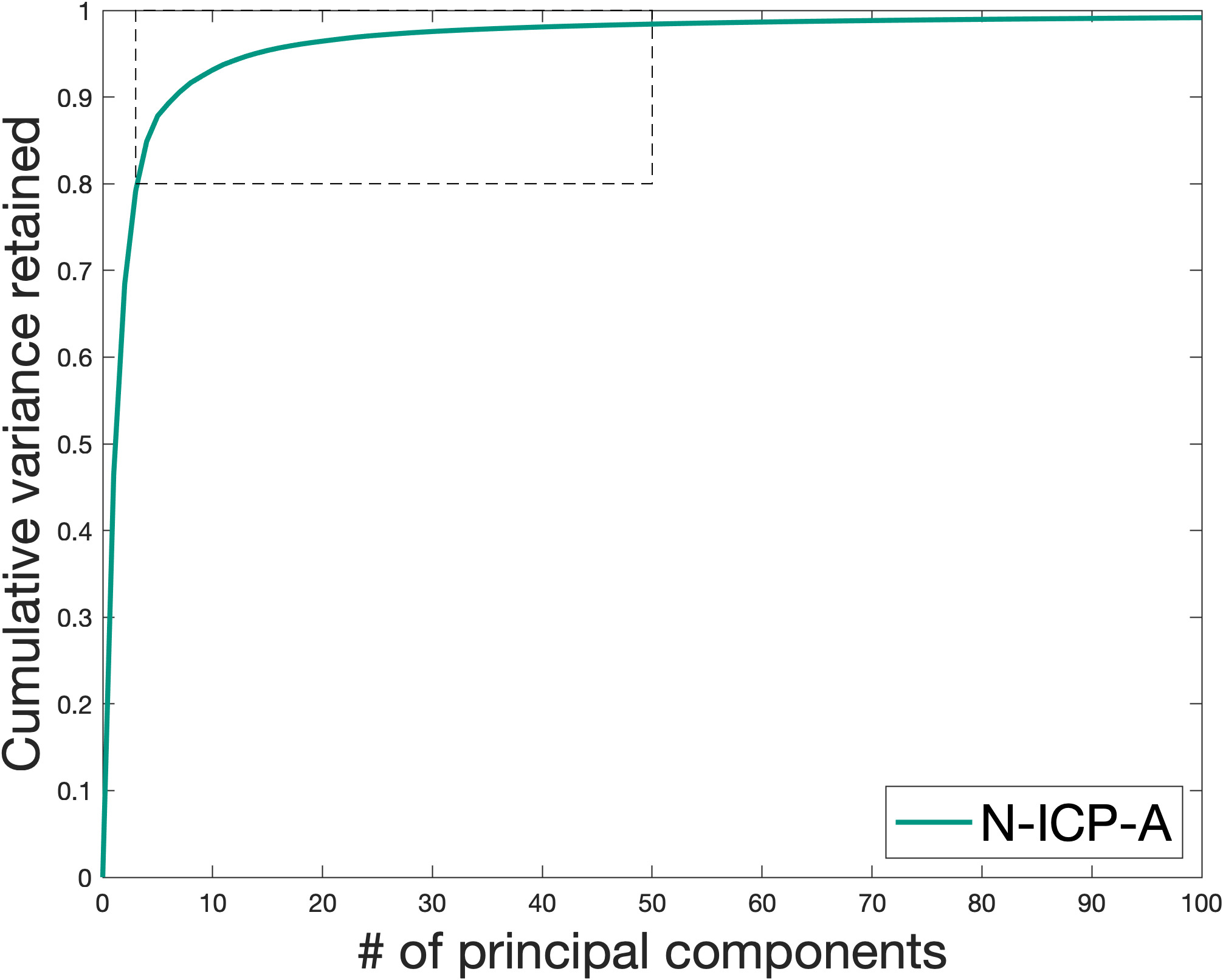}
    \includegraphics[height=\compheightIcpa\textwidth]{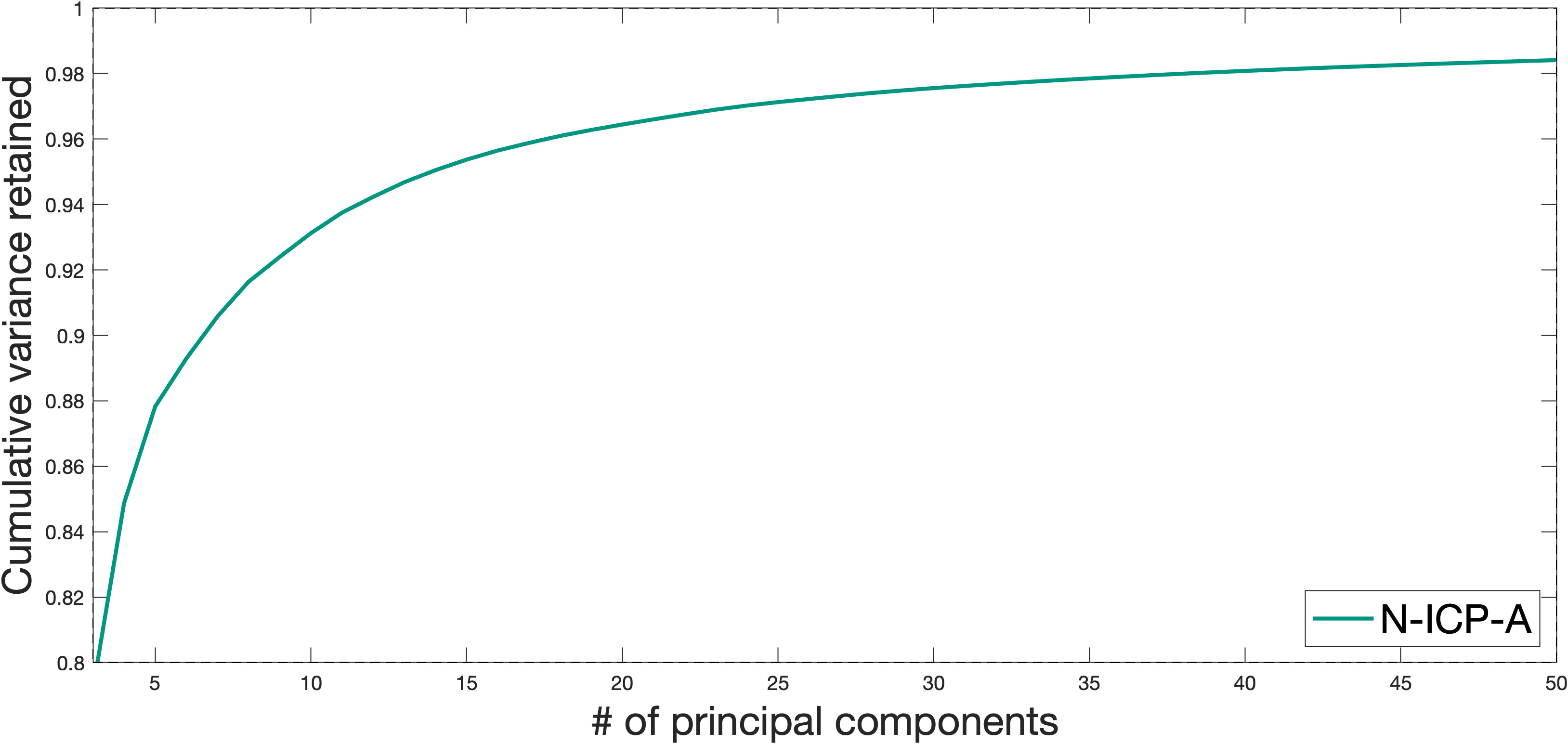}
    \newline
    \hfill
    \includegraphics[width=0.49\textwidth]{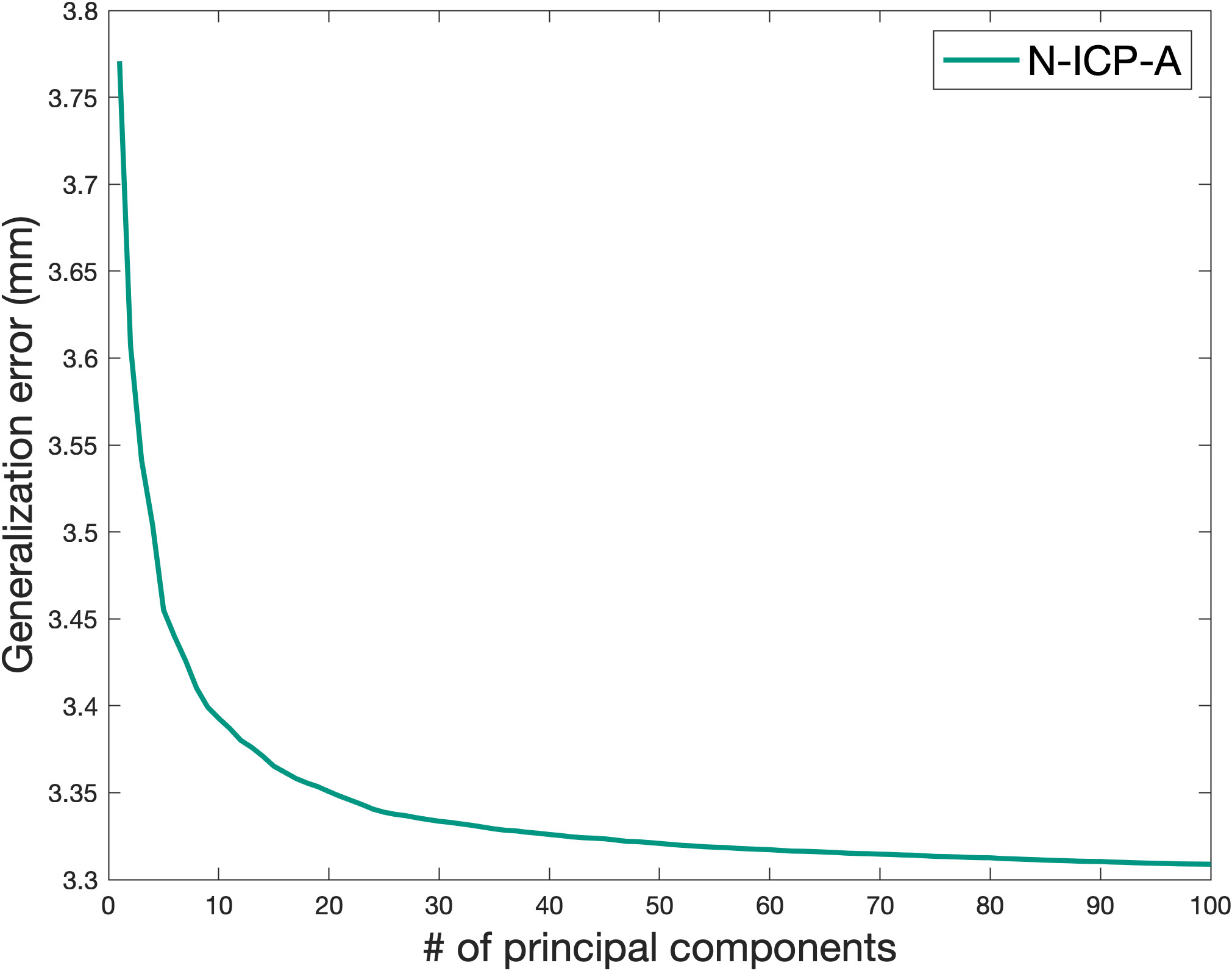}
    \hfill
    \includegraphics[width=0.49\textwidth]{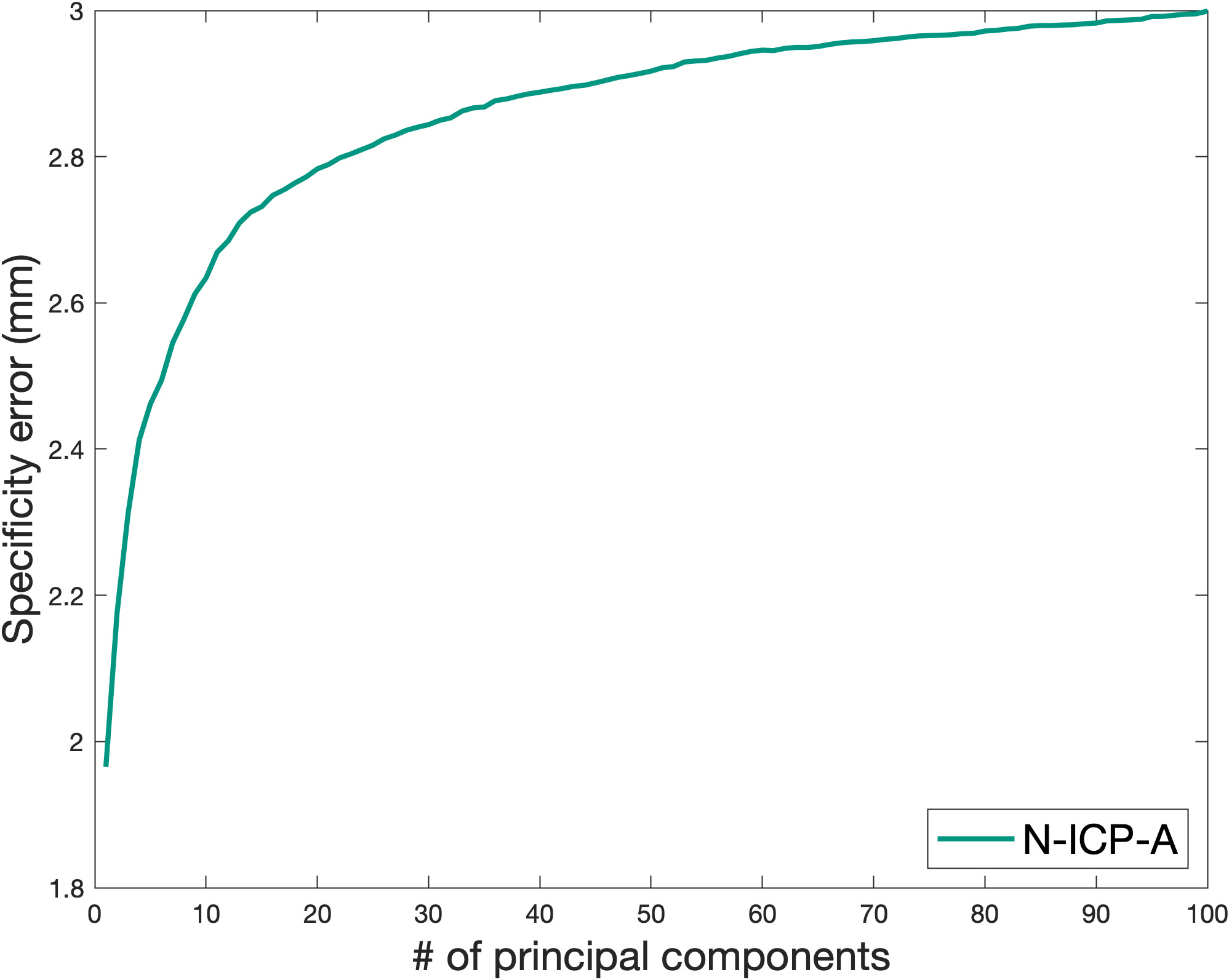}
    \hfill
    \caption{Compactness, generalization, and specificity of the final model as a function of the number 
    of principal components. Top left: Compactness, top right: Zoom-In. Higher 
    is better. Bottom left: Generalization error. Bottom right: Specificity 
    error. Lower is better.}
    \label{fig:icpaEvaluation}
\end{figure}

We performed a qualitative evaluation of the shape model eigenmodes and
submodel mean shapes on the \ac{nicpa}-model
in~\Cref{fig:principalComponents}. A change of the first eigenmode of the full
\ac{ssm} represented a change primarilly in size. For the second mode, we
observed an elongated head shape characteristic of sagittal suture fusion. In
positive direction, we observed a triangular-shaped forehead typically
associated with metopic suture fusion. The third direction showed head
asymmetry resembling left and right plagiocephaly patients present in the
control group. 

\begin{figure}[htbp]
    \centering
    \includegraphics[width=0.5\textwidth]{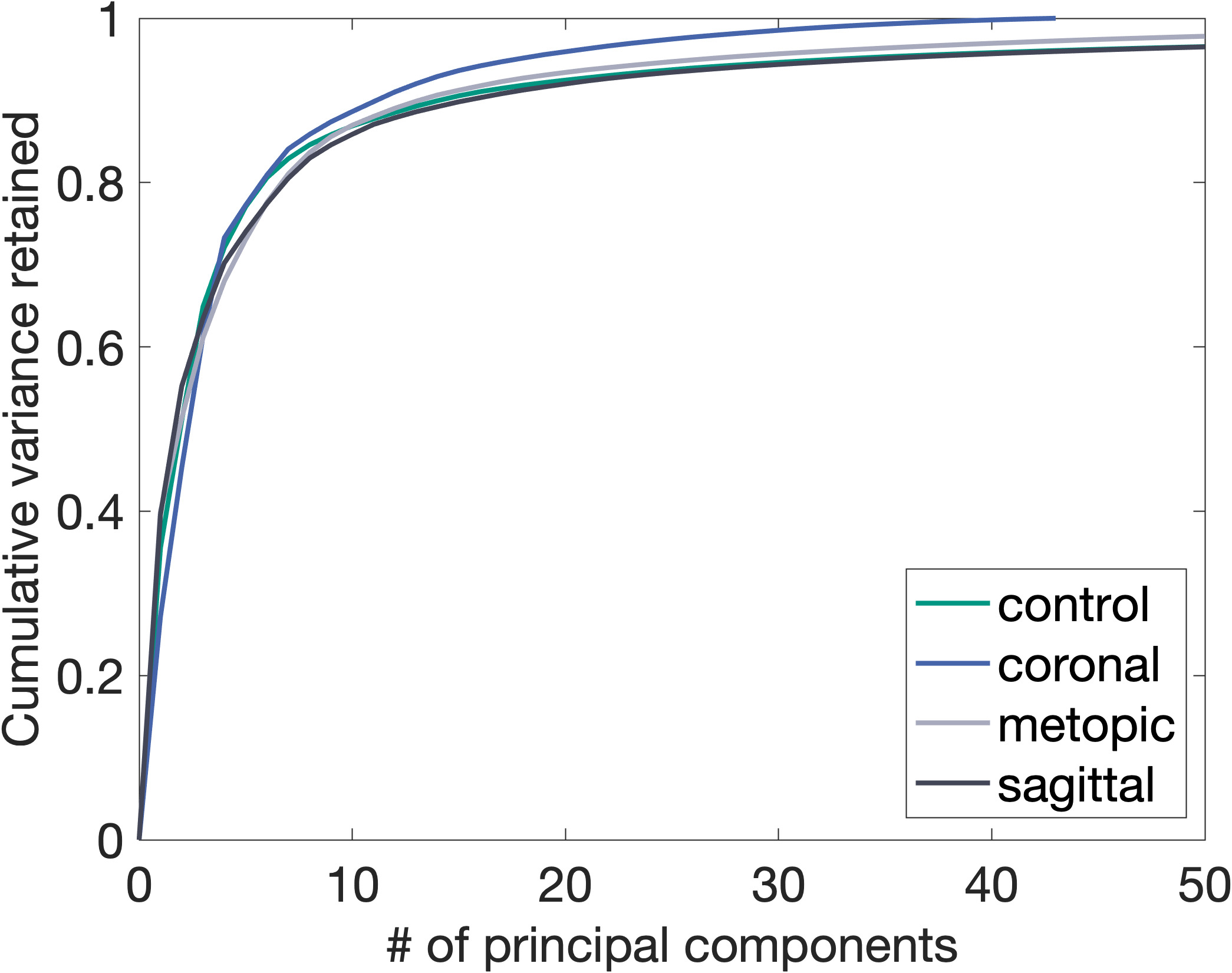}
    \newline
    \centering
    \includegraphics[width=0.5\textwidth]{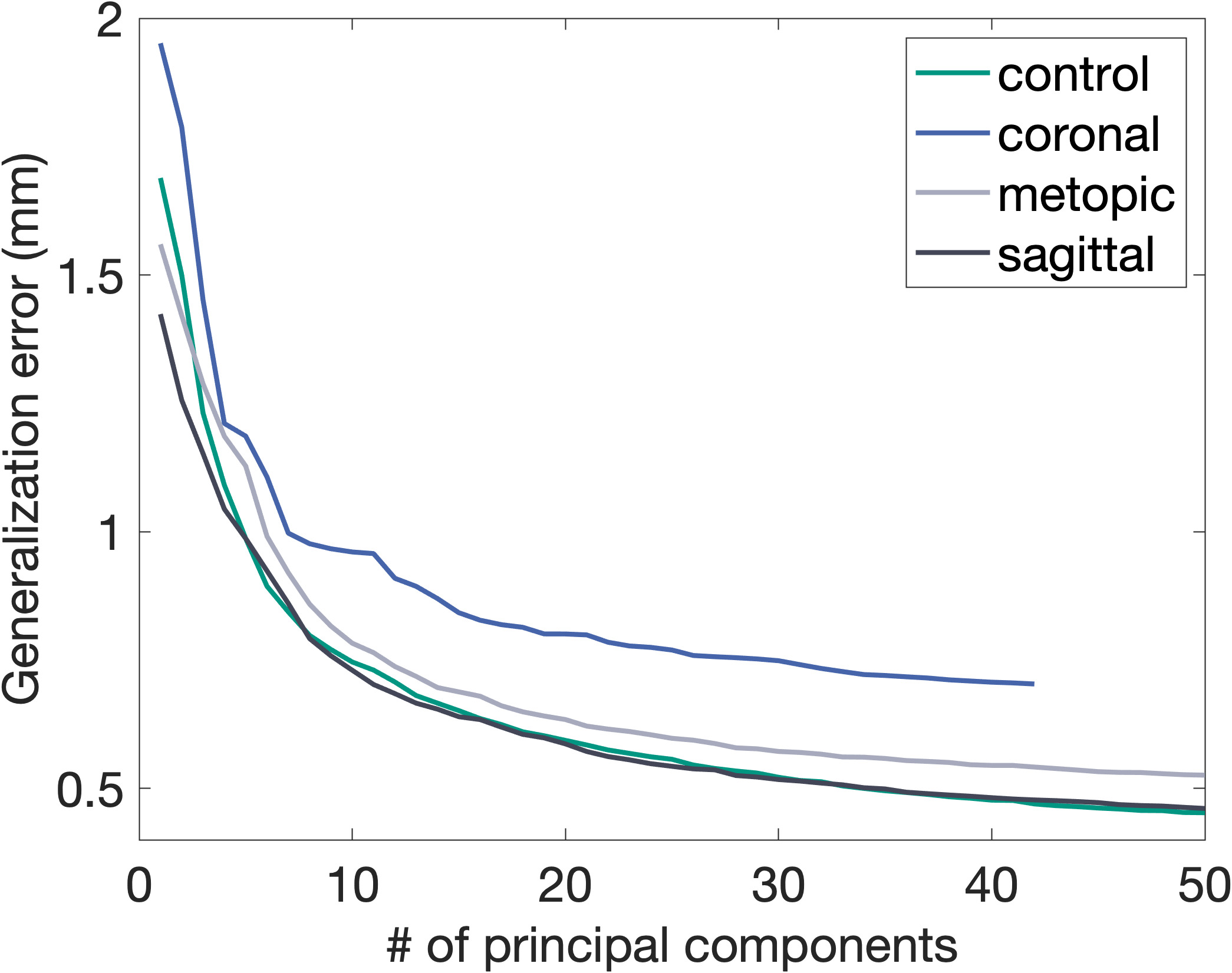}
    \newline
    \centering
    \includegraphics[width=0.5\textwidth]{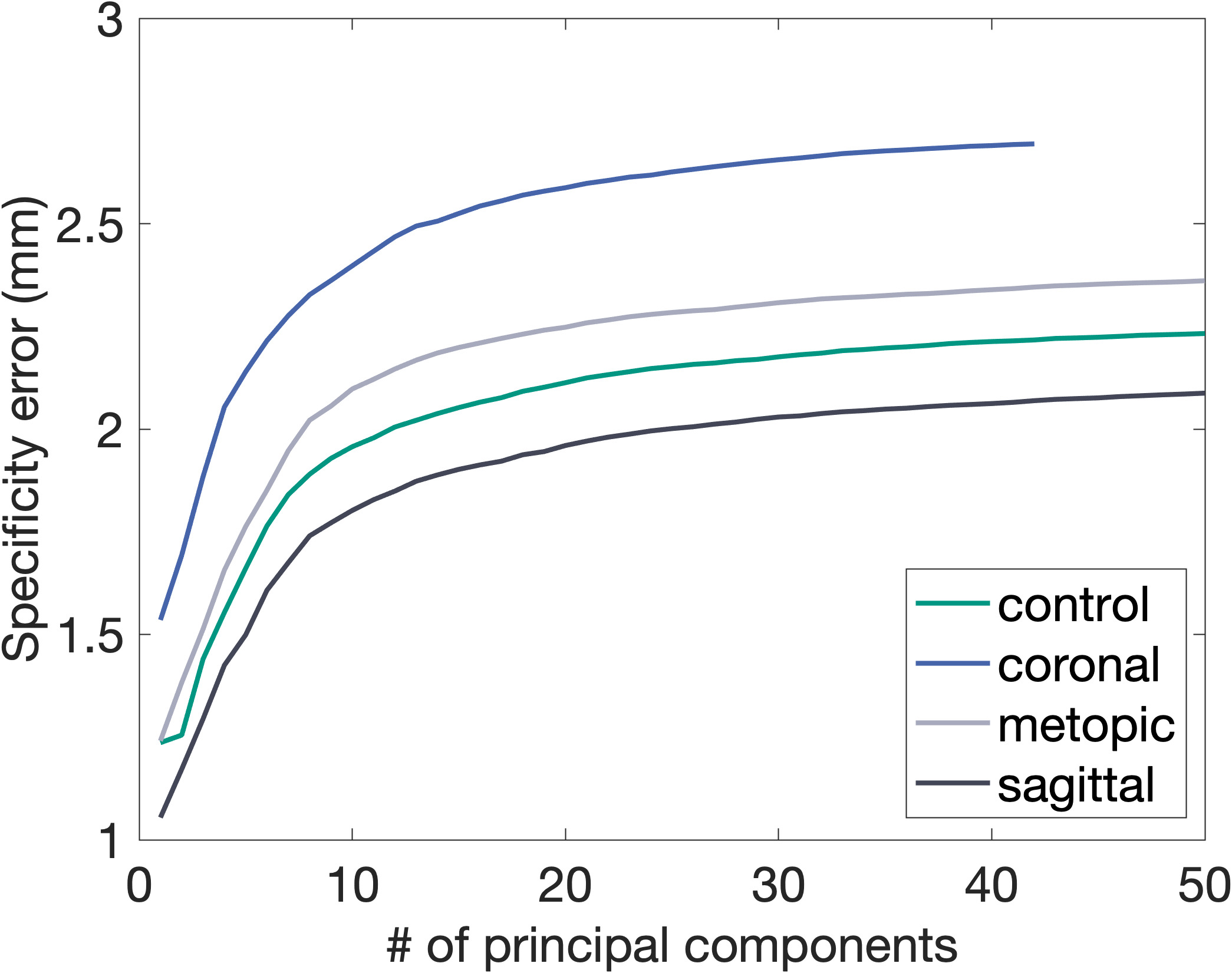}
    \newline
    \caption{Shape model metrics for the control submodel and the
    pathology-specific submodels. From left to right: Compactness,
    generalization, and specificity.}
    \label{fig:submodels}
\end{figure}

In~\Cref{fig:submodels} the resulting submodels are evaluated in terms of
compactness, generalization, and specificity. However, quantitative
comparisons between the submodels are invalid as the they differ in sample
size. Visually, the mean shapes of each pathological submodel show the typical
characteristics of each disease~(\Cref{fig:meanShapes}).

\newcommand{\msscale}{0.38}
\begin{figure}[htb]
    \begin{subfigure}[b]{0.24\textwidth}
    \centering
    \includegraphics[scale=\msscale]{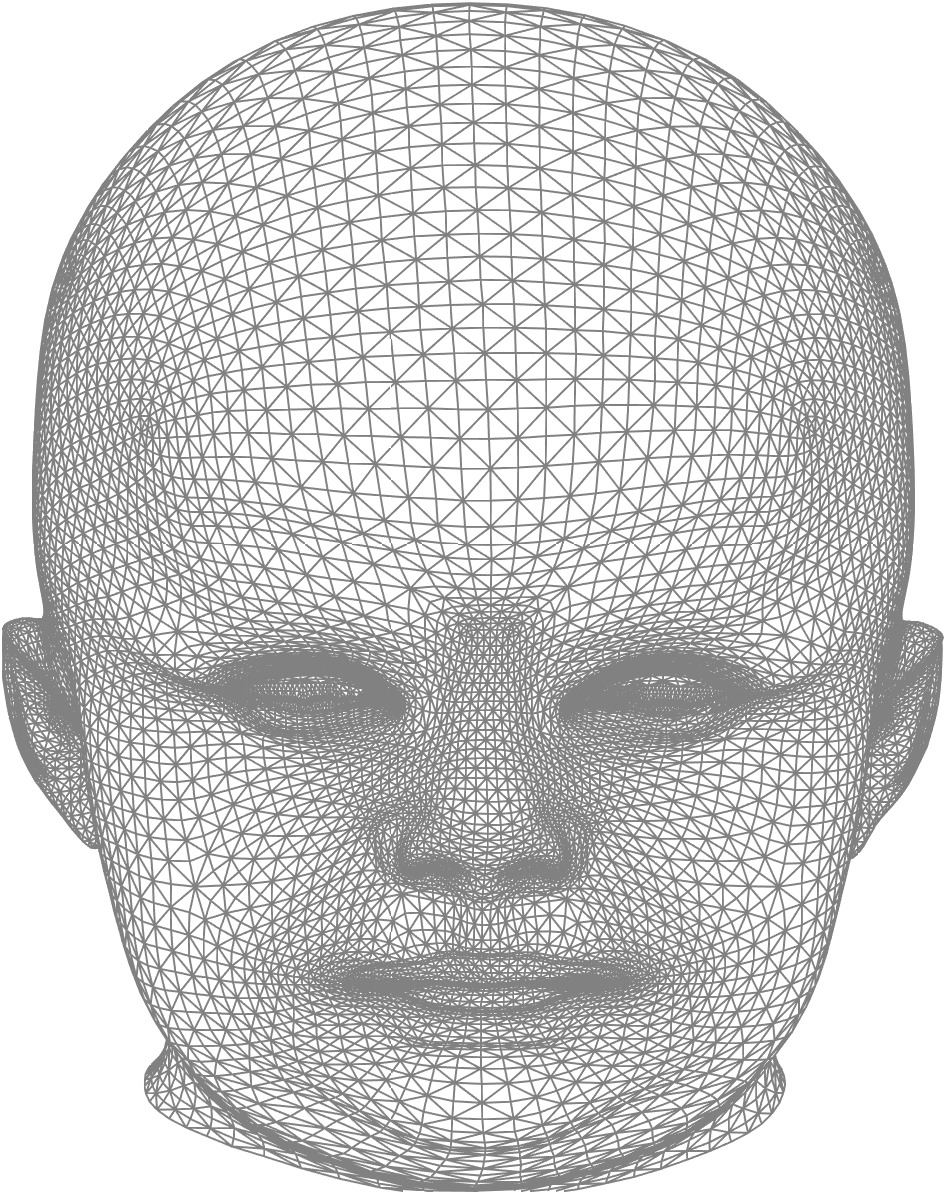}
    \centering
    \includegraphics[scale=\msscale]{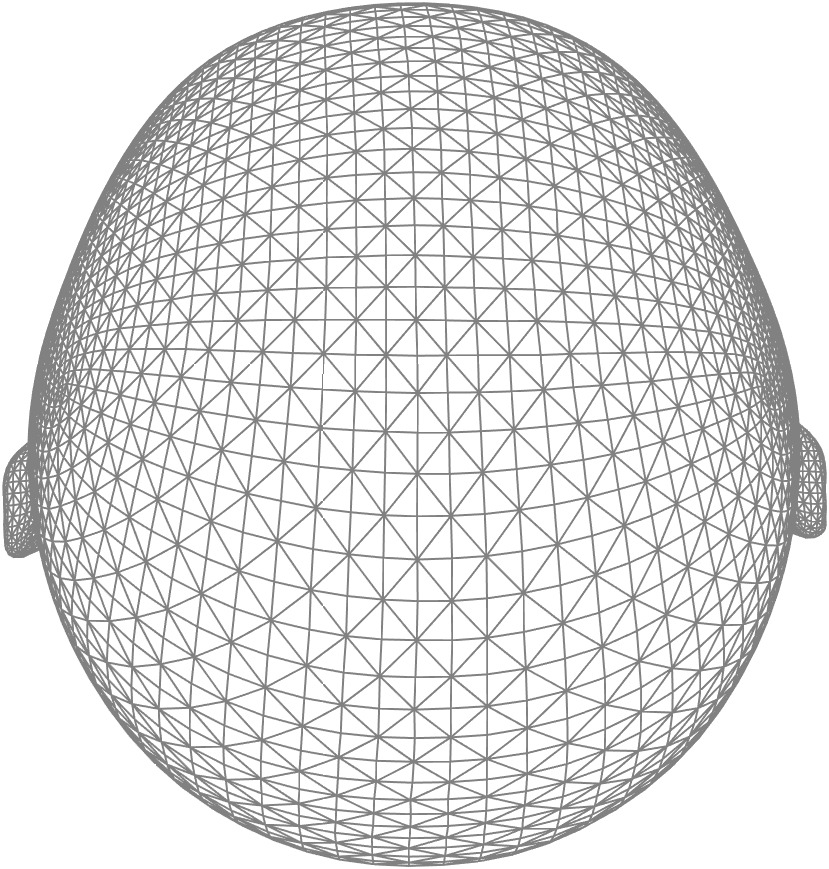}
    \caption*{\textsf{Control model}}
    \label{fig:meanHealthy}
    \vspace{0.5cm}
    \end{subfigure}
    \hfill
    \begin{subfigure}[b]{0.24\textwidth}
    \centering
    \includegraphics[scale=\msscale]{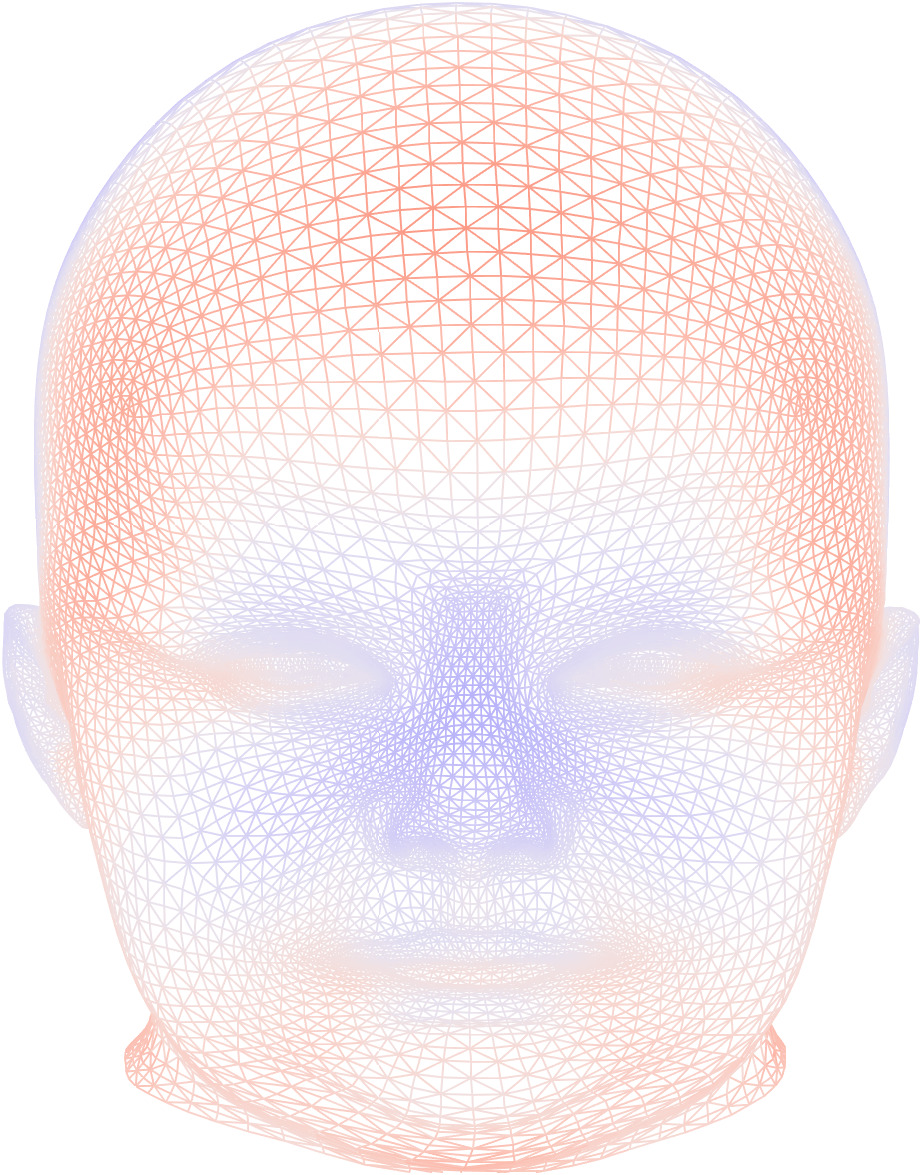}
    \centering
    \includegraphics[scale=\msscale]{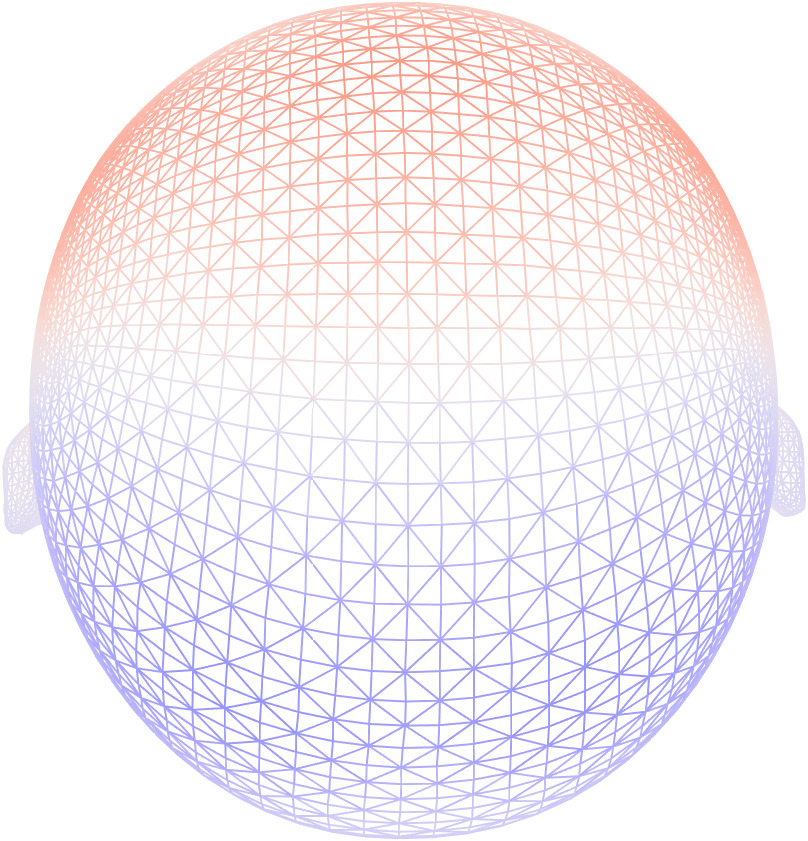}
    \caption*{\textsf{Coronal fusion model}}
    \label{fig:meancoro}
    \vspace{0.5cm}
    \end{subfigure}
    \hfill
    \begin{subfigure}[b]{0.24\textwidth}
    \centering
    \includegraphics[scale=\msscale]{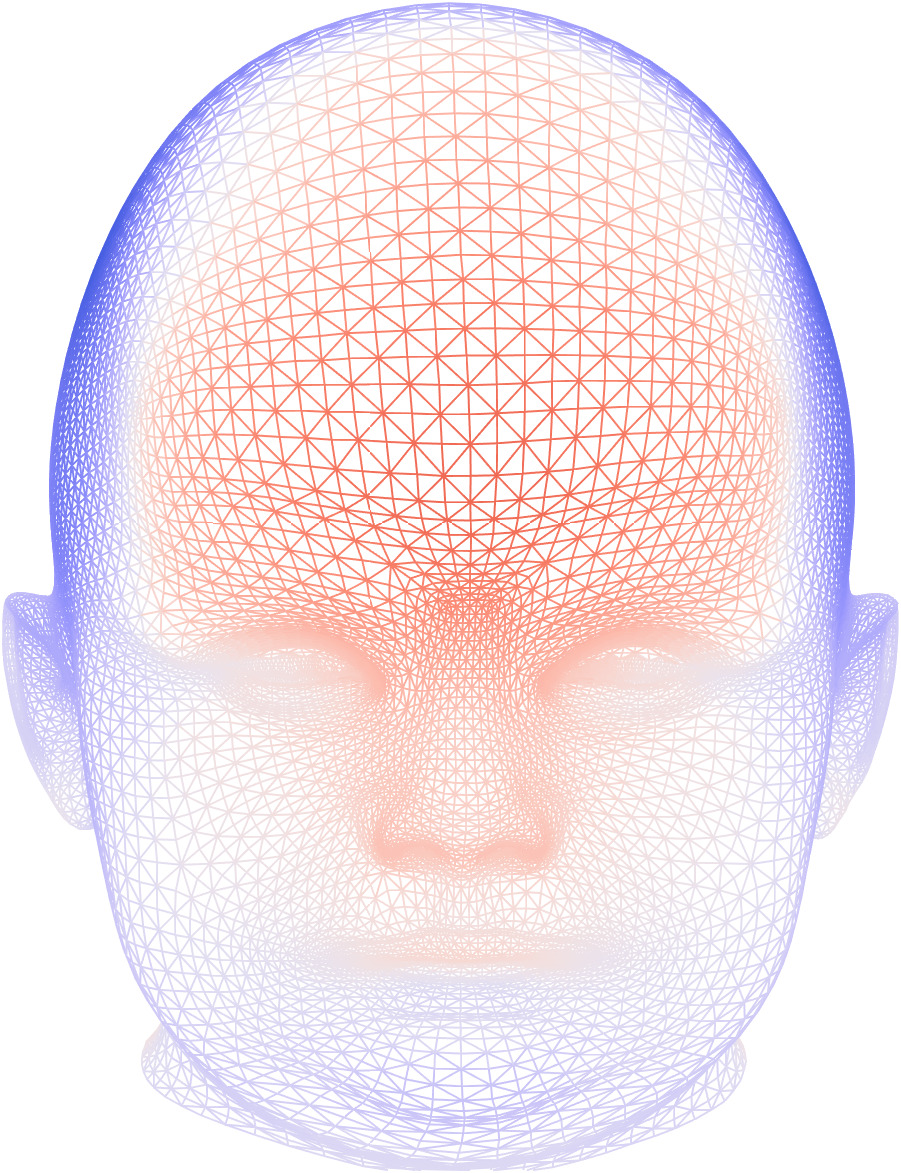}
    \centering
    \includegraphics[scale=\msscale]{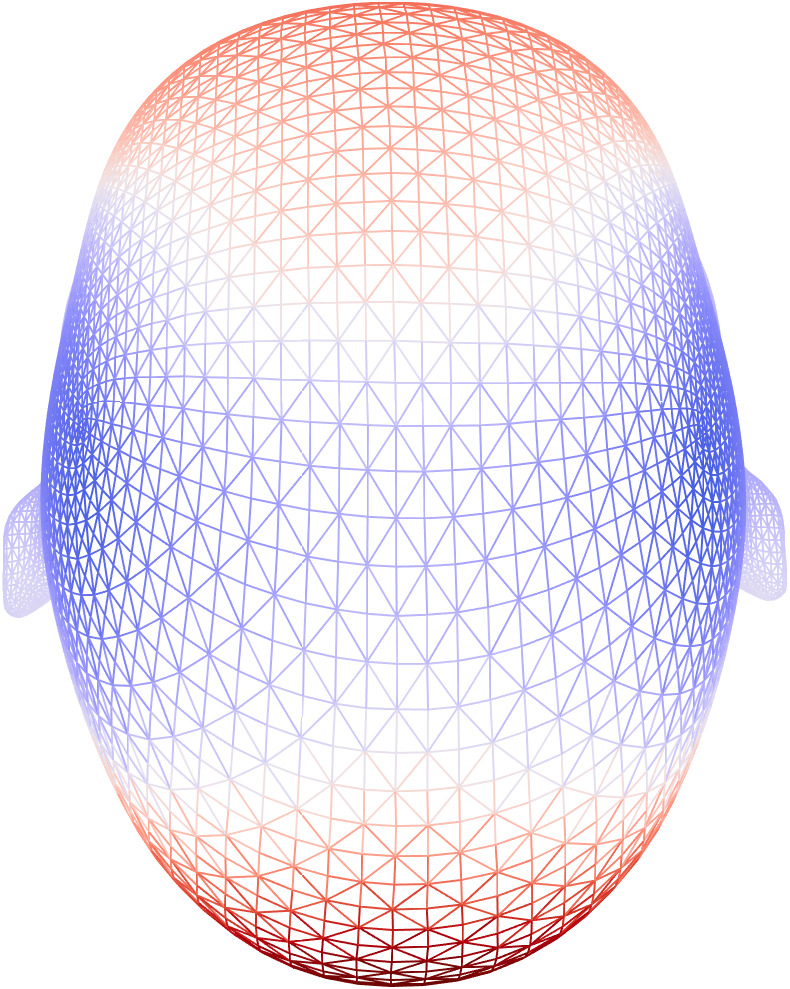}
    \caption*{\textsf{Sagittal fusion model}}
    \label{fig:meanscapho}
    \vspace{0.5cm}
    \end{subfigure}
    \hfill
    \begin{subfigure}[b]{0.24\textwidth}
    \centering
    \includegraphics[scale=\msscale]{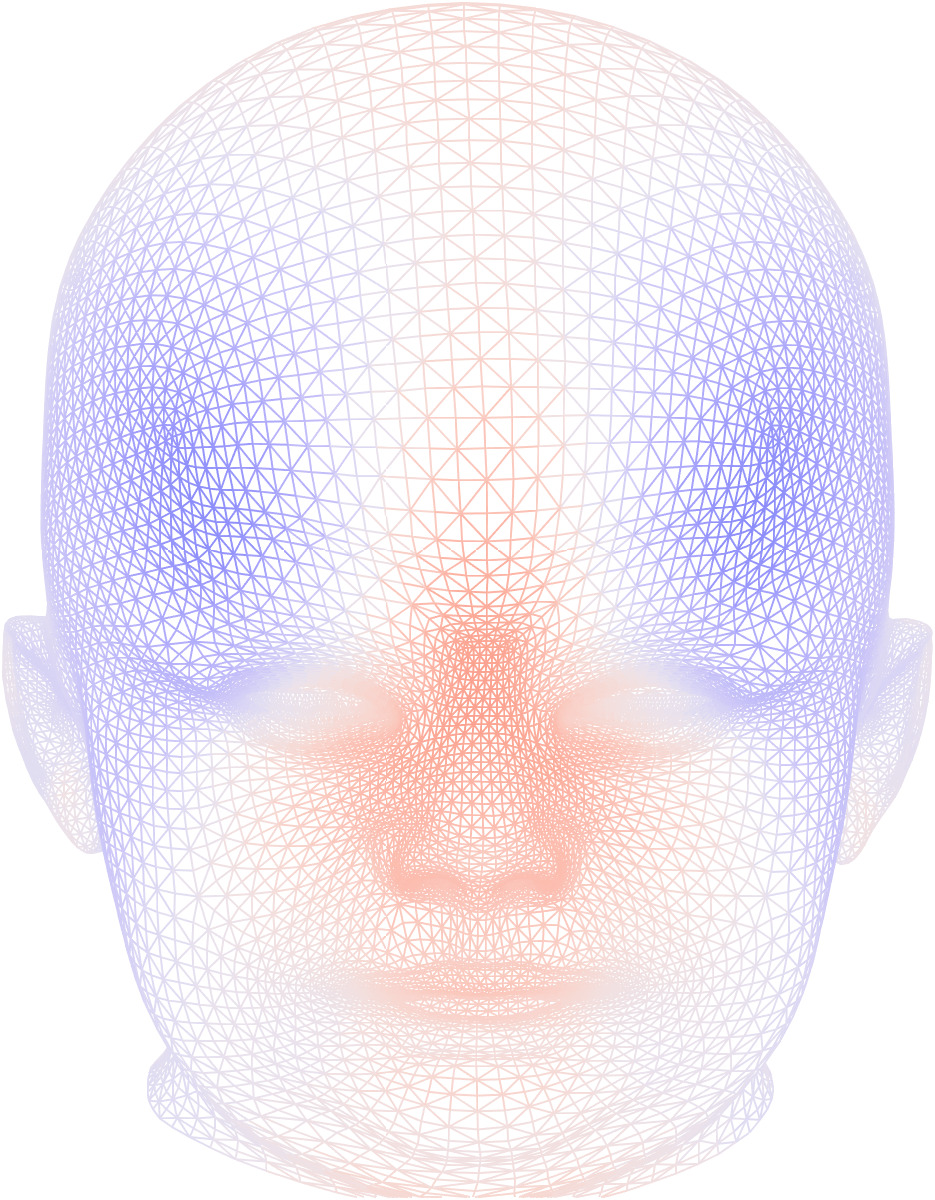}
    \centering
    \includegraphics[scale=\msscale]{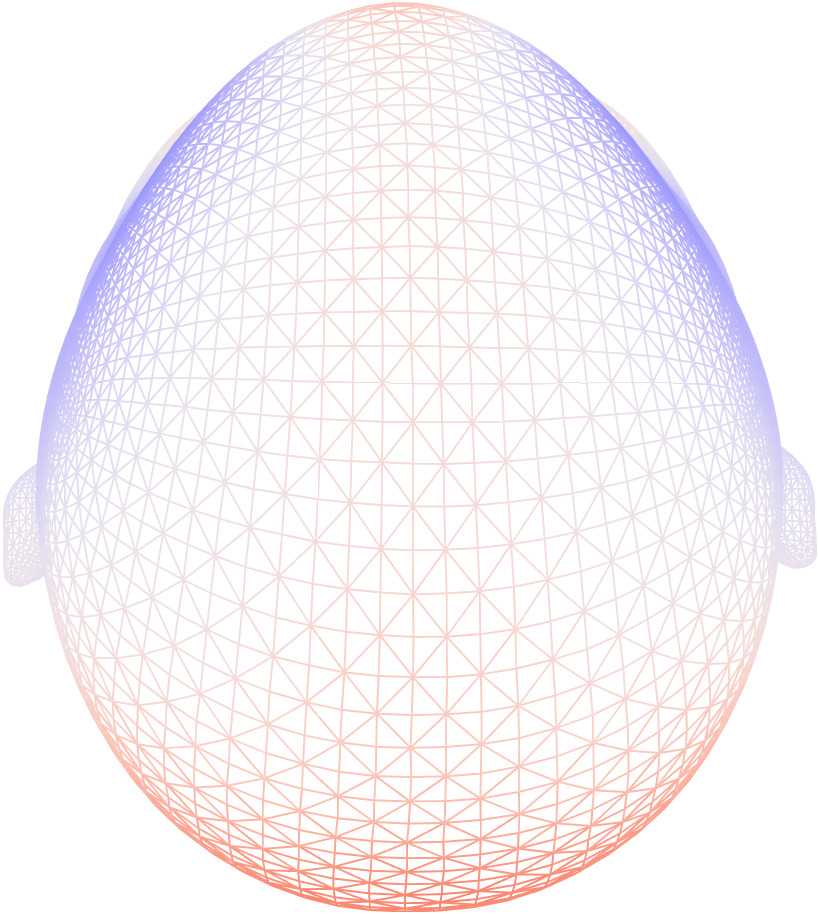}
    \caption*{\textsf{Metopic fusion model}}
    \label{fig:meantrigo}
    \vspace{0.5cm}
    \end{subfigure}
    \hfill
    \begin{subfigure}[t]{\textwidth}
    \centering
    \includegraphics[width=0.4\textwidth]{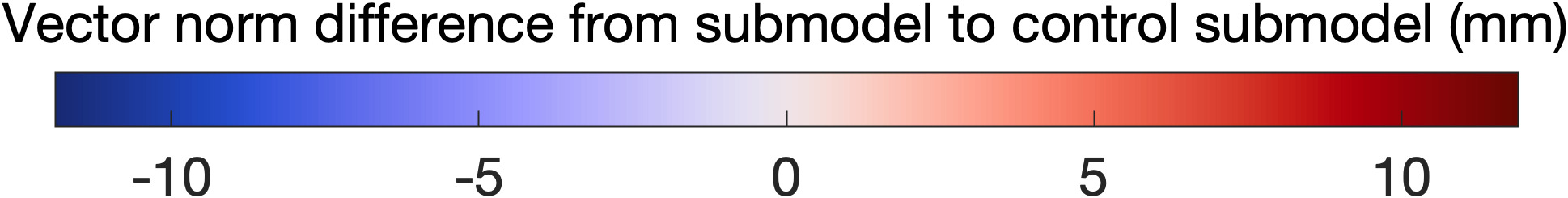}
    \end{subfigure}
    \hfill
    \caption{Mean shapes of pathology-specific submodels, front and top view.
    From left to right: control model, coronal suture fusion model, sagittal
    suture fusion model, and metopic suture fusion model. Colorbar indicates
    vector norm difference between principal component shape and mean shape
    (gray).}
    \label{fig:meanShapes}
\end{figure}

\subsection{Publicly available shape model}\label{sec:publiclyAvailableModel}

We provide the \ac{nicpa}-based \ac{ssm}, a texture model, triangular mesh
mappings, the class-specific submodels, and 100 instances of each model
sampled from a Gaussian distribution. The models are publicly available
online~\cite{schaufelberger2021a} under the Creative Commons license
CC-BY-NC~4.0. For the models, we included 95\% -- 98\% of the total variance.
As we used the Liverpool-York child head model as a basis for the initial
template for correspondence establishment, the statistical information of both
shape and texture of the models can easily be combined.
\Cref{tab:publiclyAvailable} provides information on how many components we
included in each model.

\begin{table}[htbp]
\caption{Number of principal components included in the publicly available
shape model data under Creative Commons license CC-BY-NC~4.0.}\label{tab:publiclyAvailable}
\centering
\begin{tabularx}{0.8\textwidth}{hr}
\toprule
Model & Included principal components\\
\midrule
\midrule
Full shape model & 100 \\
Texture model & 100 \\
Control model & 30 \\
Sagittal model & 30 \\
Metopic model & 25 \\
Coronal model & 15 \\
\bottomrule
\end{tabularx}
\end{table}

\subsection{Shape model applications}\label{subsec:applications}

We illustrate two applications of our \ac{ssm}. First, we changed the head
of a scaphocephaly patient toward the control group. \cite{blanz1999}
presented an approach to change an attribute (such as gender or weight) using
linear regression. As the pathologies in our model can be interpreted as such
attributes, we changed the pathology of our samples:

\begin{equation}
    \mathbf{\alpha}_\mathrm{ID,control} = \mathbf{\alpha}_\mathrm{ID} +
    \mathbf{\alpha}_\mathrm{\mu,control} - \mathbf{\alpha}_\mathrm{\mu,sagittal}
\end{equation}

We present the resulting pathology change in \Cref{fig:counselling}. This
approach can be useful in clinical settings for patient counseling.

\newcommand{\pathochangescale}{0.25}
\begin{figure}[htb]
    \begin{subfigure}[c]{0.45\textwidth}
    \centering
    \includegraphics[scale=\pathochangescale]{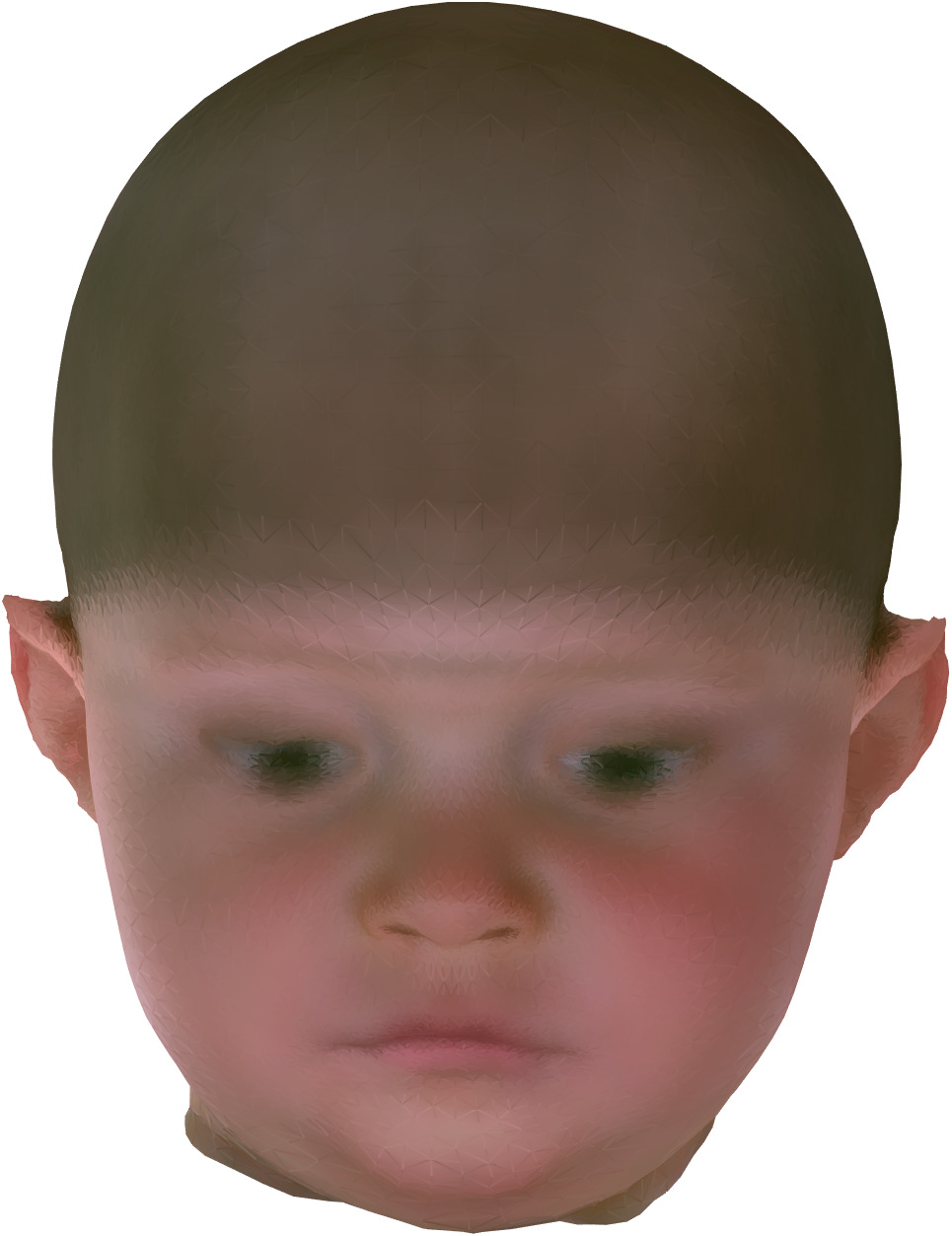}
    \quad
    \includegraphics[scale=\pathochangescale]{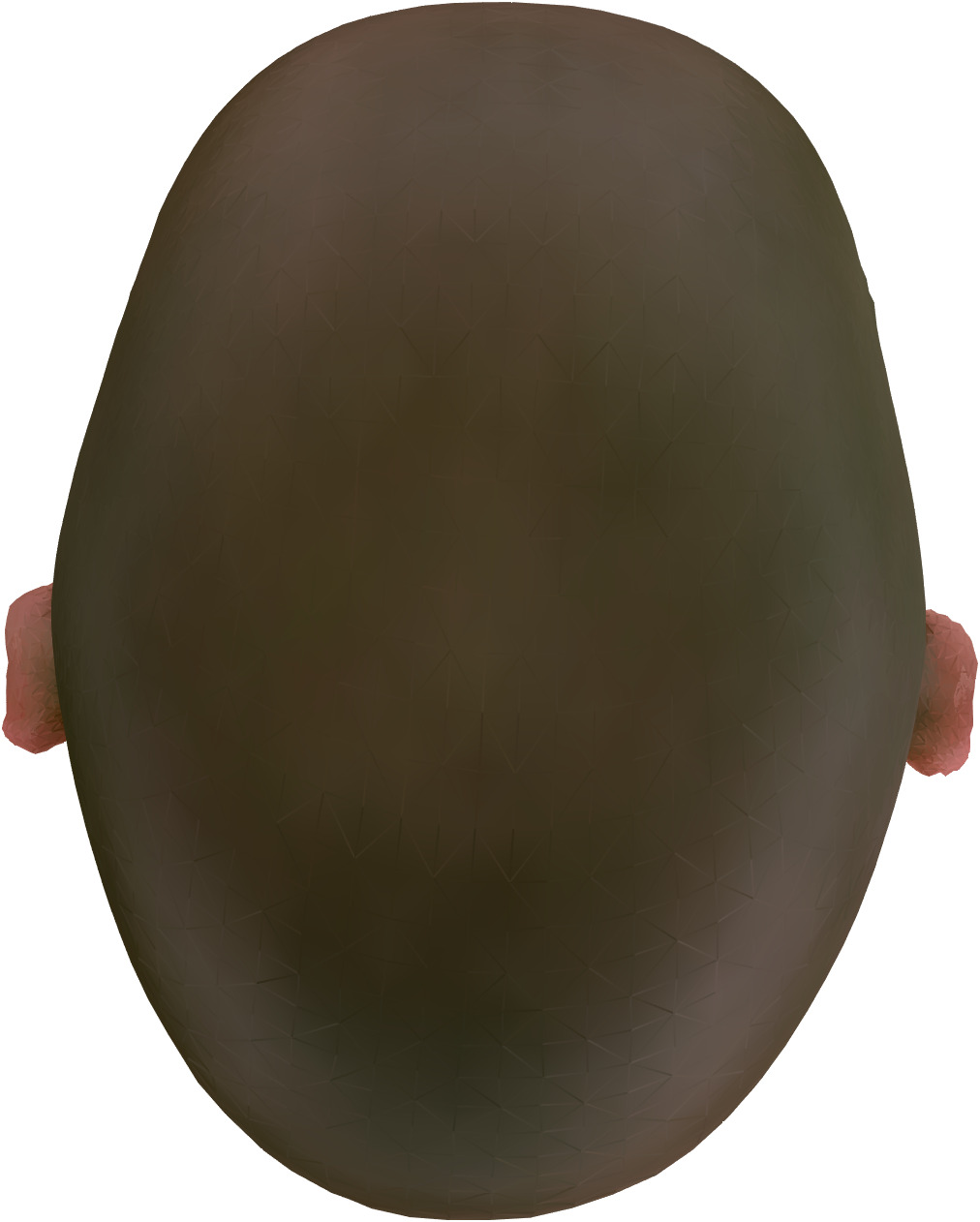}
    \caption*{\textsf{Scaphocephaly patient (sagittal suture fusion) with elongated
    head shape}}
    \label{fig:changeScapho}
    \vspace{0.5cm}
    \end{subfigure}
    \hfill
    \begin{subfigure}[c]{0.45\textwidth}
    \centering
    \includegraphics[scale=\pathochangescale]{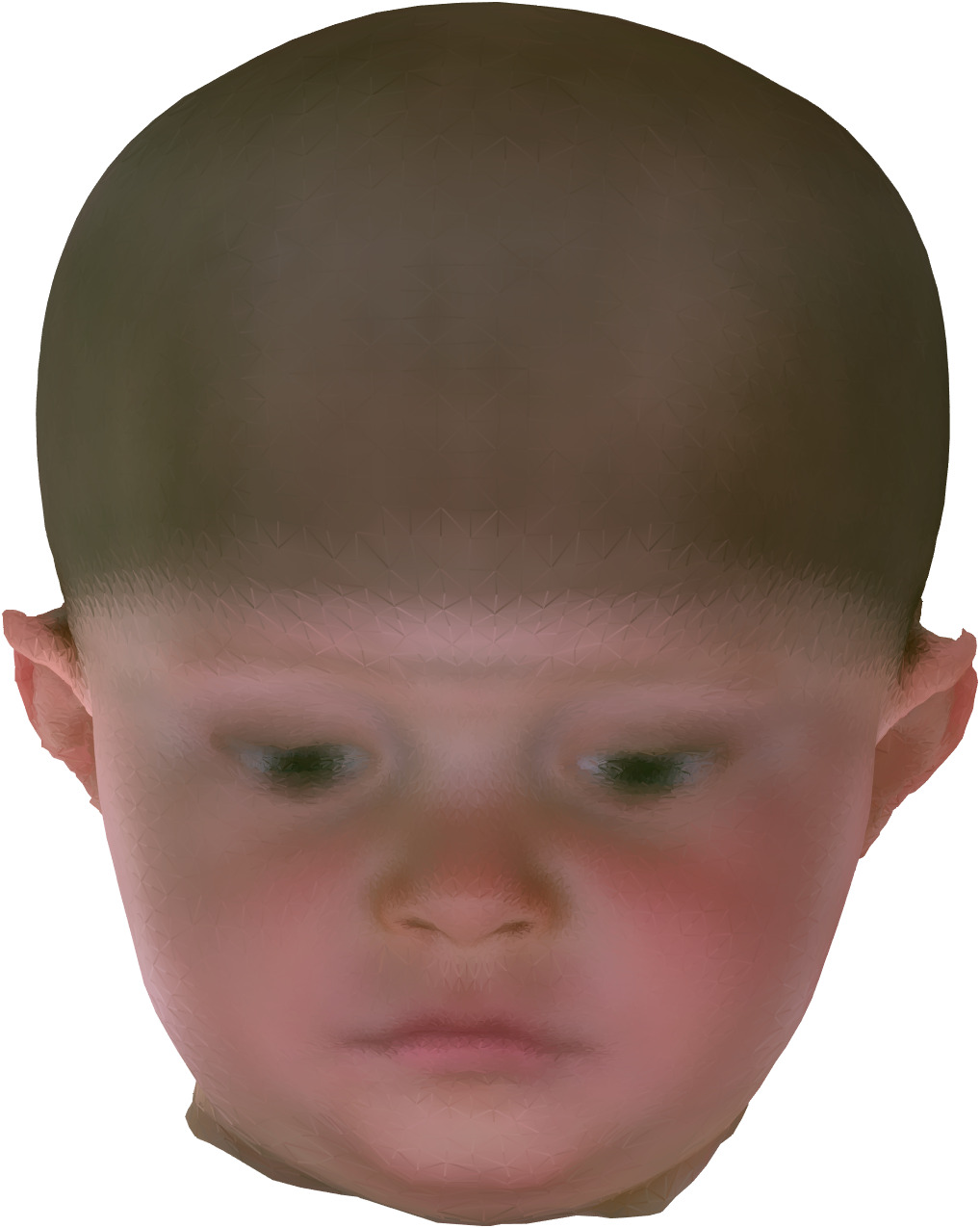}
    \quad
    \includegraphics[scale=\pathochangescale]{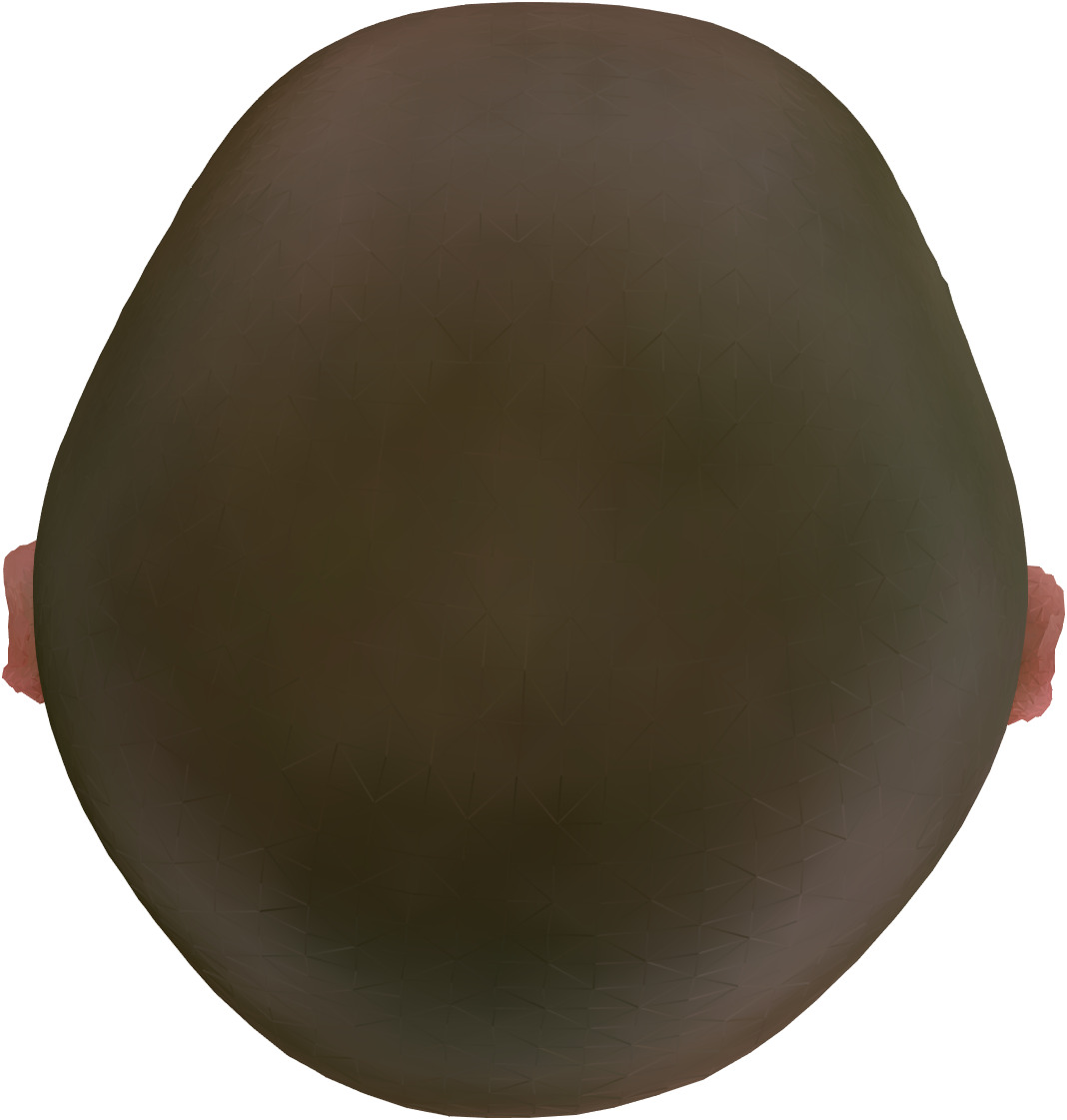}
    \caption*{\textsf{Synthetic representation of the patient without scaphocephaly}}
    \label{fig:changeHealthy}
    \vspace{0.5cm}
    \end{subfigure}
    \caption{Patient pathology assessment using pathology change.  Left: the
    original head shape of the scaphocephaly patient. Right: the patient's
    head with removed pathology using our full \ac{ssm}.}
    \label{fig:counselling}
\end{figure}

Second, we sampled random instances from our shape model while keeping the
points on the cranium fixed. Modeling the remaining flexibility on a shape
which is held partially fixed can be described as a constrained generalized
eigenvalue problem~\cite{albrecht2008}. This approach can be applied to
create a synthetic database for machine learning applications and is depicted
in~\Cref{fig:flexmodes}. Alternatively, synthetic samples with predefined
pathology can also be created using a \ac{psm}~\cite{albrecht2013} or by
simply using the pathology-specific submodel. 

\newcommand{\fmodescale}{0.24}
\begin{figure}[htb]
    \begin{subfigure}[c]{0.32\textwidth}
    \includegraphics[scale=\fmodescale]{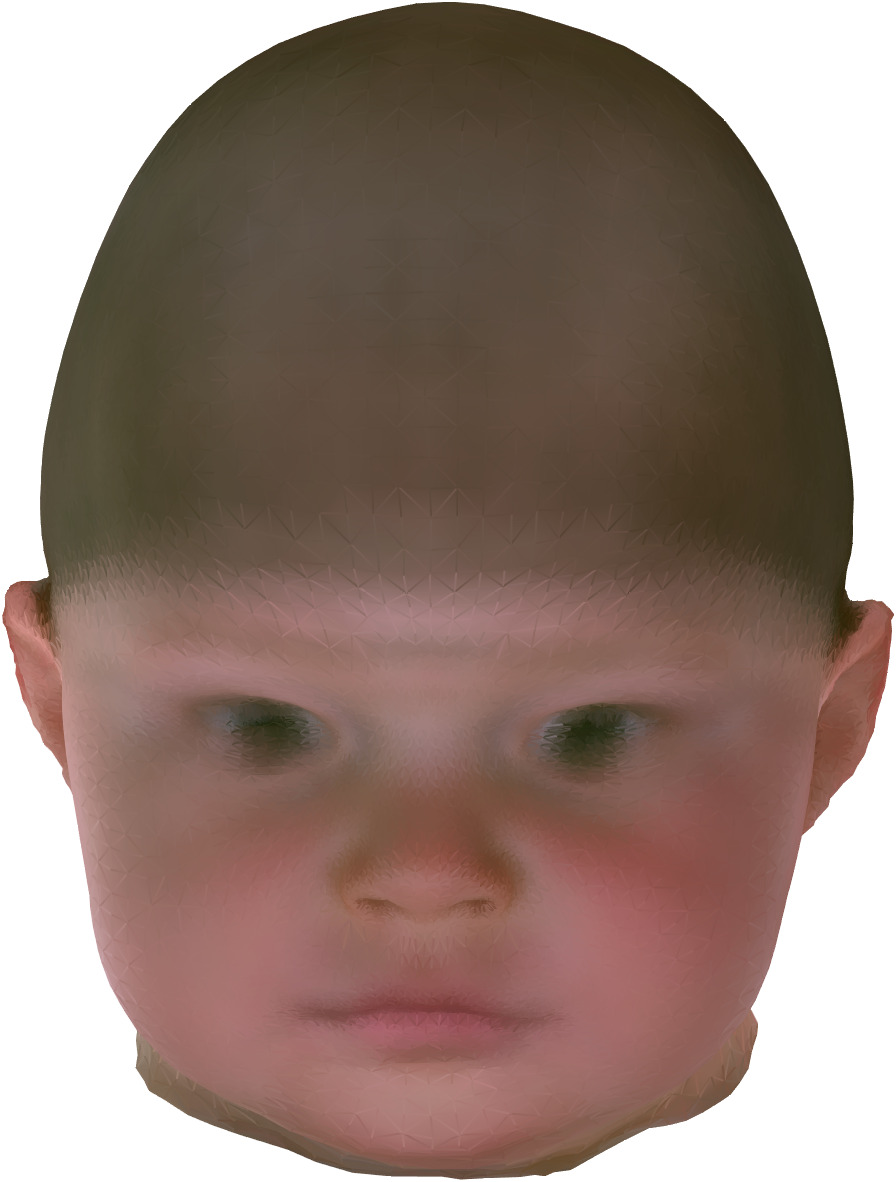}
    \includegraphics[scale=\fmodescale]{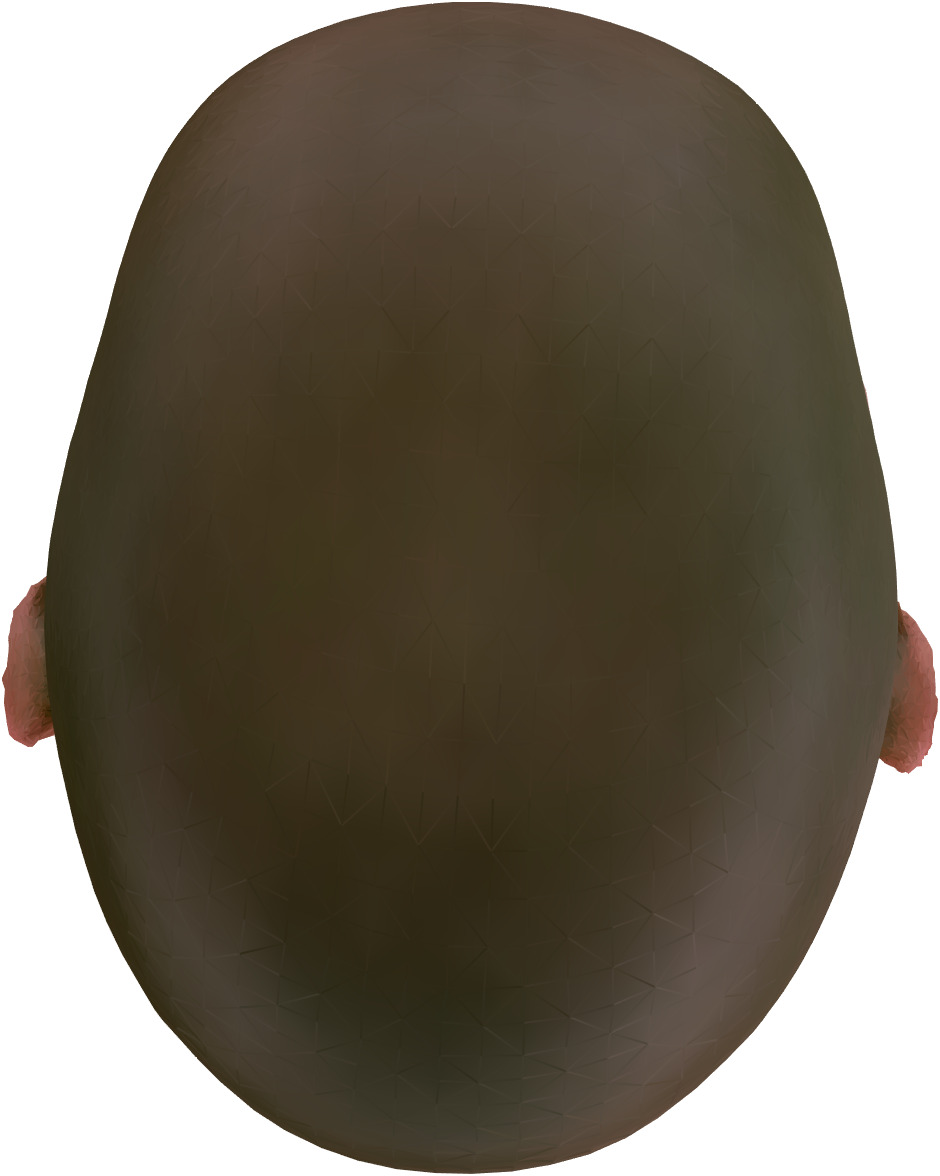}
    \caption*{\textsf{First flexibility mode.}}
    \end{subfigure}
    \hfill
    \begin{subfigure}[c]{0.32\textwidth}
    \includegraphics[scale=\fmodescale]{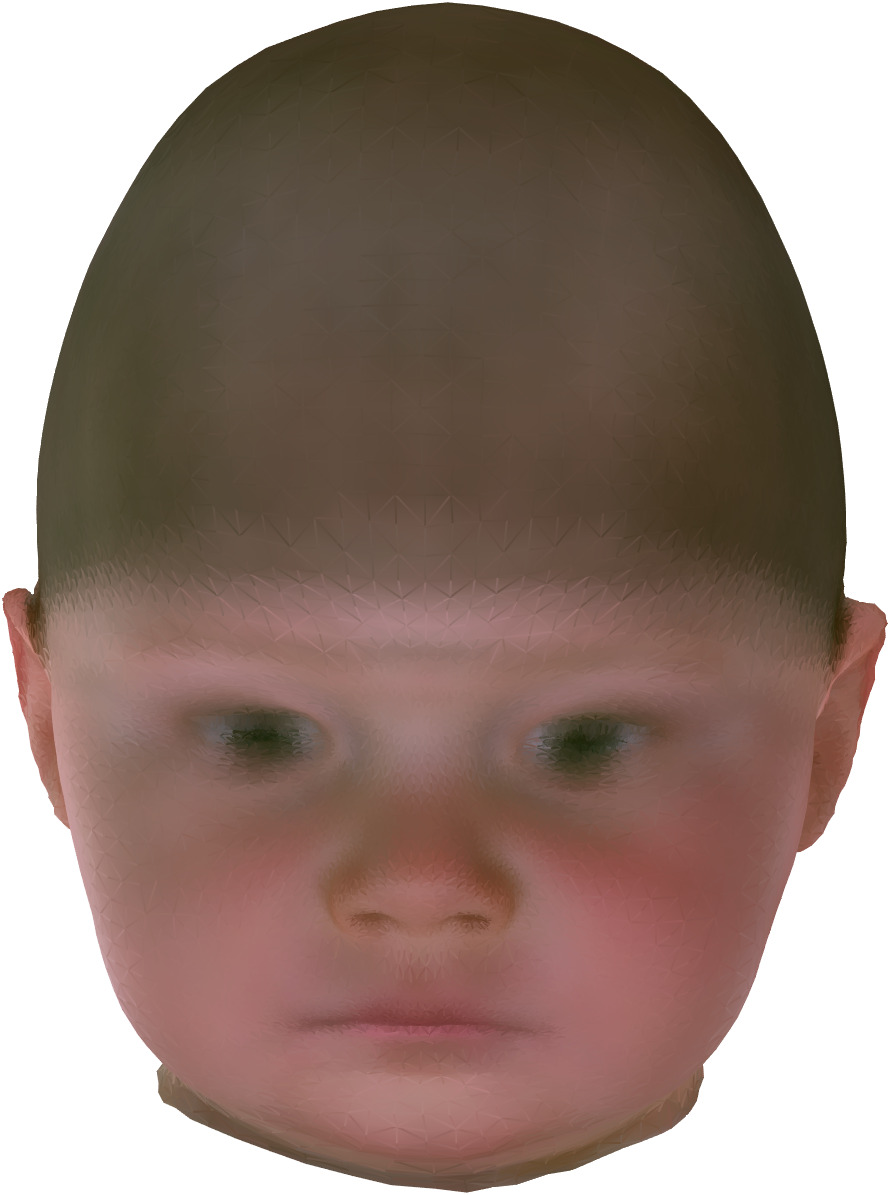}
    \includegraphics[scale=\fmodescale]{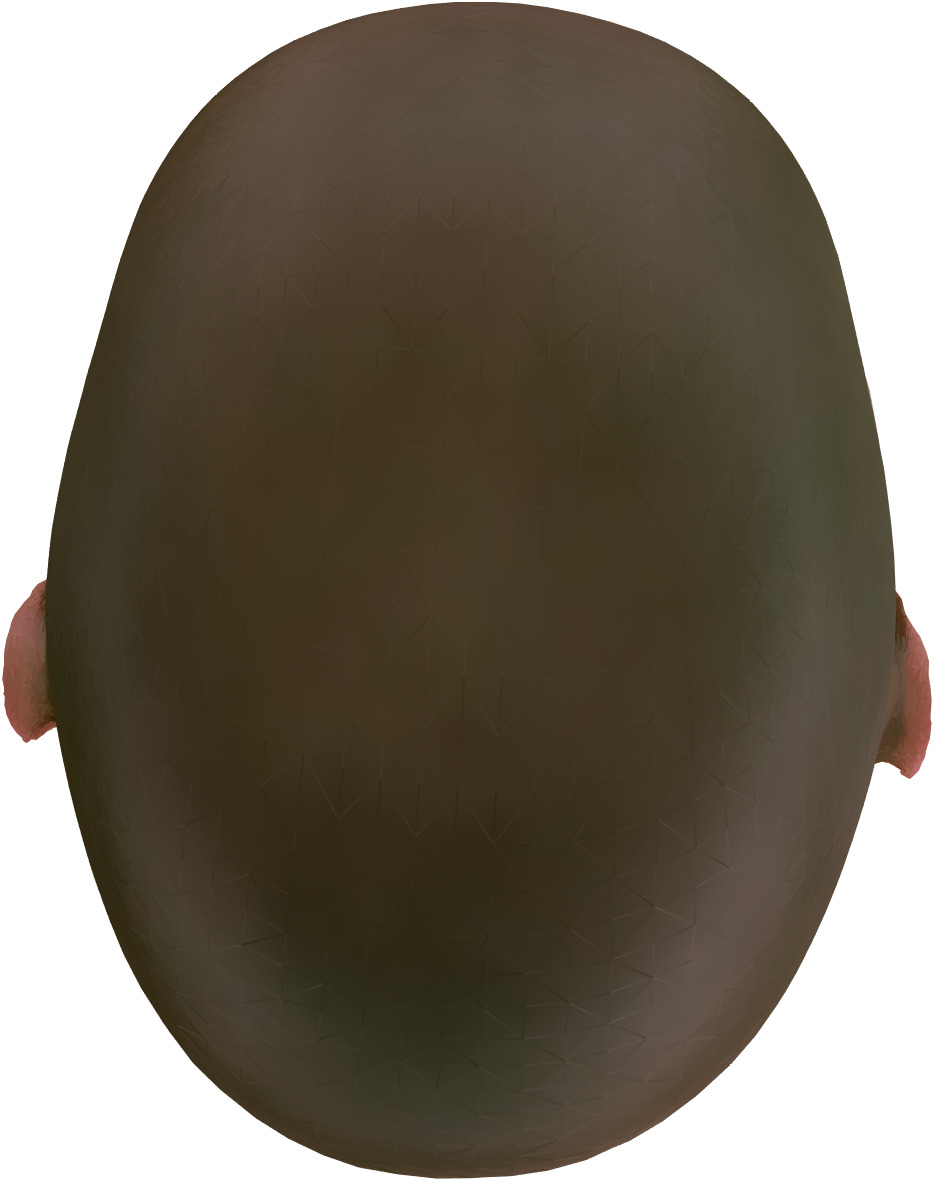}
    \caption*{\textsf{Second flexibility mode.}}
    \end{subfigure}
    \hfill
    \begin{subfigure}[c]{0.32\textwidth}
    \includegraphics[scale=\fmodescale]{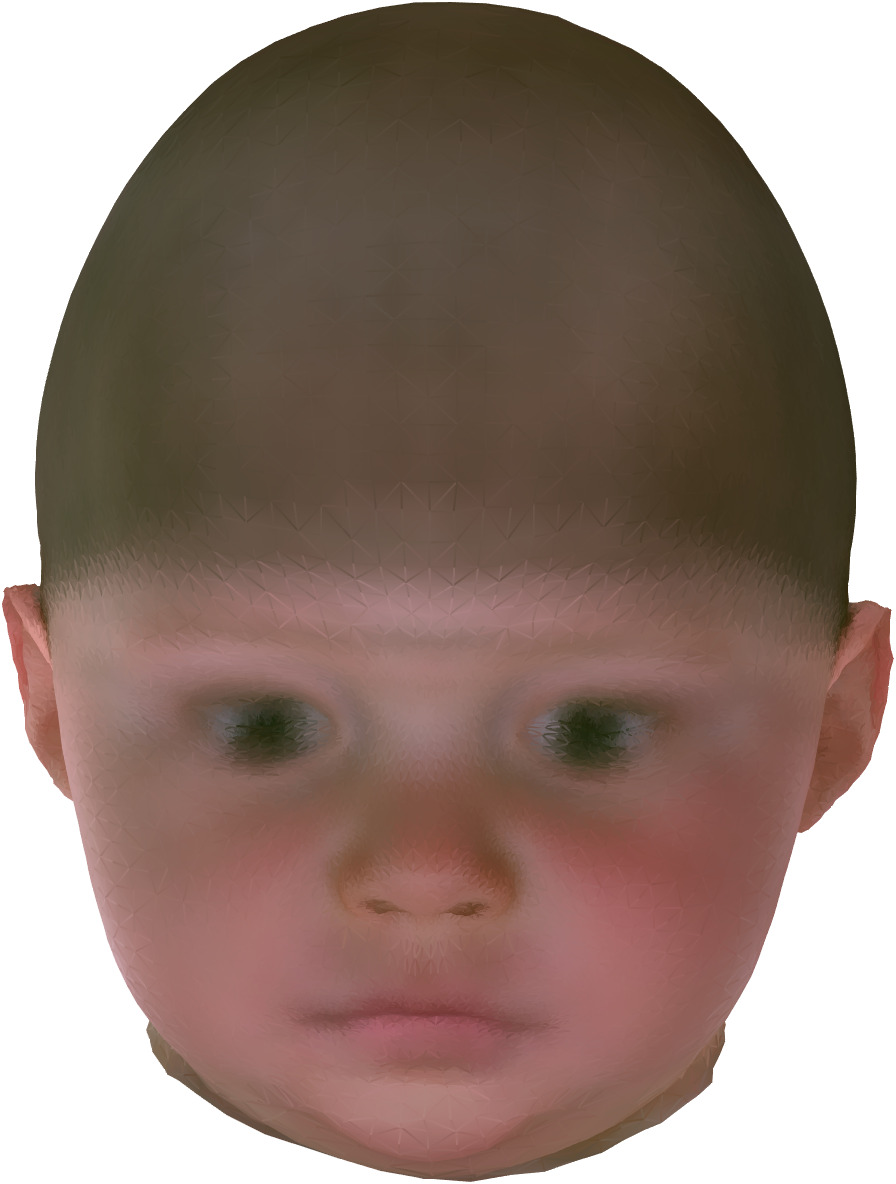}
    \includegraphics[scale=\fmodescale]{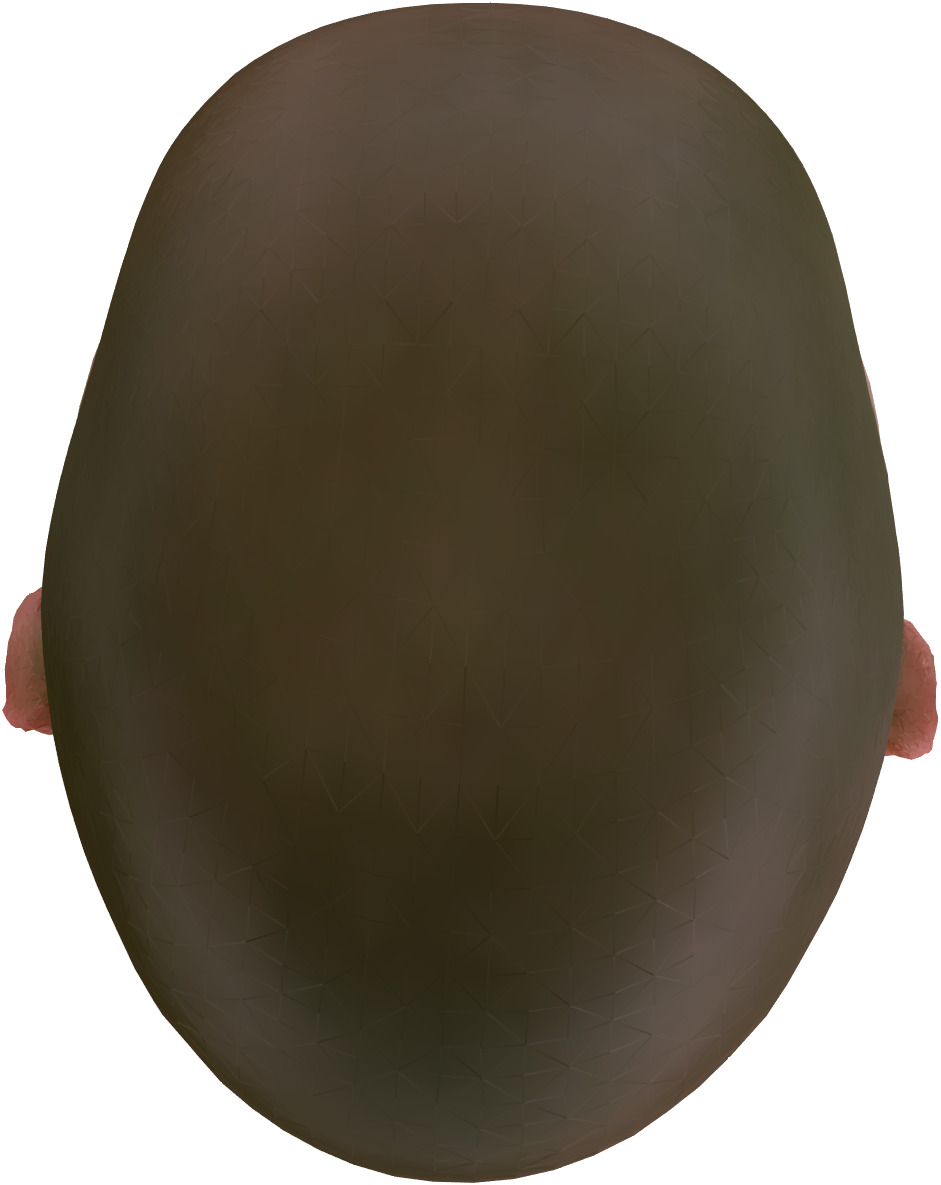}
    \caption*{\textsf{Third flexibility mode.}}
    \end{subfigure}
    \caption{The first three flexibility modes with fixed cranium, applied to a
    synthetic scaphocephaly patient. Changes are minimal for the cranium and
    maximal for the face, neck, and ears.}
    \label{fig:flexmodes}
\end{figure}

\subsection{Classification results}\label{subsec:classificationEvalutation}

\begin{figure}[ht]
\centering
\includegraphics[width=\textwidth]{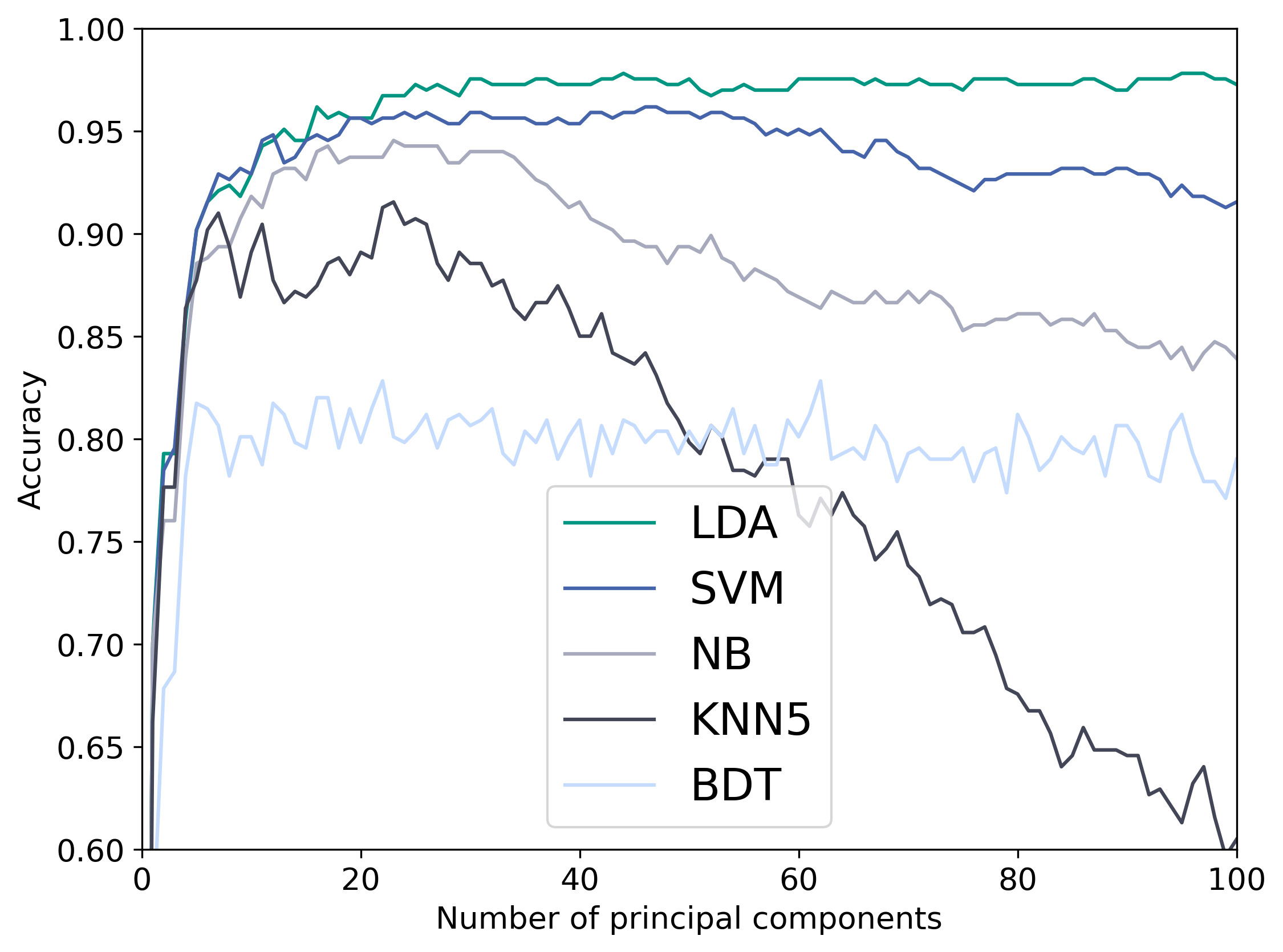}
\caption{Accuracy as a function of the number of principal components used for the
\acf{nicpa} classifier.}\label{fig:accOverPC}
\end{figure}

We tested the five classifiers on the cranial models of each of the four
morphing methods. In this section, we only show results for \ac{nicpa}. A
comparison of the classifiers on the other morphing methods can be found
in~\Cref{ap:morphingClassification}. \Cref{fig:accOverPC}~shows the accuracy
over the used number of principal components on the \ac{nicpa} approach.
\Ac{lda}, \ac{svm}, and \ac{nb} outperformed \ac{knn} and \ac{bdt}. A
reduction in accuracy with adding more principal components could be observed
for \ac{nb} and \ac{knn} and less pronounced for the \ac{svm}.

For the optimal classification setup (\ac{lda} with 44 components) we show
confusion matrix, per-class sensitivities, per-class specificities, and g-mean
in~\Cref{tab:classification}. The classifier yielded optimal per-class
specificities for the pathological cases and the per-class sensitivity for the
controls and metopic cases. The remaining per-class sensitivities for the
pathological cases were between 0.773 and 0.973 while the per-class specificity
for the control group was 0.958. The g-mean resulted in 0.931 and the total
accuracy in 0.978.

\begin{table}[ht]
\caption{Confusion matrix, sensitivity and specificity using \acf{lda},
\acf{nicpa}, and the optimal number of components (44). Con = control, Cor =
coronal, Sag = sagittal, Met = metopic.}\label{tab:classification}
\centering
\small
\begin{tabularx}{0.9\textwidth}{XttttXX}
\toprule
True class & \mbox{Predicted class} & & & & Sensitivity & Specificity\\
\midrule
& Con & Cor & Met & Sag & & \\
\midrule
\midrule
Con & 178 & 0  & 0  & 0   & 1.000 & 0.958 \\
Cor & 5   & 17 & 0  & 0   & 0.773 & 1.000 \\
Met & 0   & 0  & 56 & 0   & 1.000 & 1.000 \\
Sag & 3   & 0  & 0  & 108 & 0.973 & 1.000 \\
\midrule
\midrule
G-mean & & & & & 0.931 & \\
\mbox{Total accuracy} & & & & & & 0.978 \\
\bottomrule
\end{tabularx}
\end{table}

\section{Discussion}\label{sec:discussion}

We presented the first publicly available craniosynostosis \ac{ssm}. It unites
statistical information of 367 subjects and their mirrored twins with and
without craniosynostosis. To date, many methods presented by various authors
rely on in-house datasets making quantitative comparisons difficult. A set of
synthetic photogrammetric head scans of our \ac{ssm} can help creating a large
patient cohort for a reproducible evaluation of methods to assess
craniosynostosis.

Our model reflects the pathologies available in our dataset: The first two
components show changes in size as well as changes related to sagittal and
metopic suture fusion. These are the two largest craniosynostosis classes in
the dataset. The third principal component is associated with head asymmetry,
resulting from non-synostotic positional plagiocephaly subjects in the control
class. The pathology-specific submodels also depict the typical head
deformities observed in clinical studies and are best used for controlled
sampling of labeled synthetic instances.

From the tested template morphing methods, none of the approaches clearly
outperformed the other ones. We considered \ac{nicpa} and \ac{icpdlbrp} to be
the two most promising methods. Our decision to use \ac{nicpa} over
\ac{icpdlbrp} for the publicly available model was motivated by the smaller
vertex-to-nearest-neighbor-distances and the higher model compactness.
Landmark errors are typically considered the gold standard, but as our model
is concerned with craniosynstosis and the landmarks are located primarily on
the face, we deemed them less important for this model. 

A comparison with other craniosynostosis-related \acp{ssm} is difficult as we
propose the first publicly available model of infants. In the medical field,
studies which used shape models~\cite{heutinck2021,lamecker2006} did not
include quantitative metrics such as landmark error, compactness,
generalization, and specificity. Thus, we compare our model to a
non-craniosynostosis-related \acp{ssm} of the full head. The most comparable
\ac{ssm} might be the Liverpool-York-Model~\cite{dai2017a,dai2020}, as it is
a full head model and also contained a submodel comprising children from 2 to
15 years. Compared with the Liverpool-York head model, our model is more
compact, has a lower specificity error, and lower nertex-to-nearest-neighbor
distances, but a higher generalization and landmark error. However a direct
comparison may not be meaningful considering that both the dataset and the
used mesh resolution are different. Our training data contains only children
younger than 1.5 years, so the total variance of our model might be smaller by
nature. However, by including a similar \ac{lbrp}-approach in our analysis,
we show that our model performs comparable to state-of-the-art models.

The second contribution of this work is a classification pipeline approach
using our \ac{ssm}. To the best of our knowledge, we tested it on the largest
dataset used for a craniosynostosis-related classification study to date.
Multiple authors~\cite{meulstee2017,heutinck2021} have shown that shape
modeling enables a quantitative analysis of the head shape with respect to
craniosynostosis. In our work, we demonstrated that \ac{ssm} can not only
quantify, but also classify head deformities. With an accuracy of 97.8\,\% on
367 subjects, our approach classifies craniosynostosis comparable to competing
methods: \cite{mendoza2014} achieved a classification accuracy of 95.7\% on
141 subjects using \ac{ct} data and \cite{deJong2020} obtained an accuracy of
99.5\,\% on 196 samples using a feedforward neural network in combination with
ray casting and stereophotographs. As each classification approach used
different datasets, it can be argued that quantitative comparisons between
different approaches might be dataset dependent. However, as we tested
multiple classifiers and multiple morphing methods, we demonstrate that our
classification approach is robust and does not rely on heavy hyperparameter
tuning. Morphing methods showed little influence in the final classification
accuracy. The choice of the classifier had a larger influence on the
classification results: \ac{lda} and \ac{svm} appeared to be the most robust
classifier with respect to the noisy components, while \ac{nb} worked well
with fewer than 40 components.

The control class of our study is assembled by the scans of children who
visited the hospital without indication to be treated surgically. This
includes patients who were diagnosed being healthy and patients who were
diagnosed having mild head deformities due to positional plagiocephaly. Thus,
the control model represents a mixed group of children and should be used with
caution when generating healthy subjects.

Using \ac{pca} assumes that the training data follows a multivariate normal
distribution. This assumption does not hold up for a head model which includes
different pathology classes. With respect to the classification, \ac{pca}
serves as a reparameterization and ultimately as a dimensionality reduction
procedure. This seems to be one of the key elements of our classification
approach. \Ac{pca} is the de-facto standard method for \ac{pdm} generation,
although some authors have proposed some alternatives for specific cases.
\Ac{ppca} has been proposed for datasets with missing data or
outliers~\cite{luthi2009}, but has higher computational costs. As we have a
regular mesh and have removed corrupt scans before establishing dense
correspondence, we employed \ac{wpca}.

Many authors proposed modifications to further improve \acp{ssm}. With respect
to template morphing, multiple correspondences can be taken into
account~\cite{liang2018}. Some possible improvements for the statistical
modeling include the use of \ac{ppca}~\cite{luthi2009}, reparameterization of
training shapes~\cite{davies2003}, the use of part-based
models~\cite{dai2019}, or combining multiple
models~\cite{ploumpis2019,ploumpis2021}. \cite{luthi2018}~presented
\acp{gpmm}, modeling deformations as Gaussian processes, which increases model
flexibility by the use of prior models, combining kernels, and operating in
the continuous domain. \cite{gerig2017}~made use of domain-specific Gaussian
process shape deformations for model building. These methods might further
improve our model.

\section{Conclusion}\label{sec:conclusion}

We presented the first head model of craniosynostosis patients and made it
publicly available. We included pathology specific submodels, ready-to-use
sampled instances of each submodel, and a texture model. Our model performs
similar to state-of-the-art head models with respect to morphing and model
metrics and captures craniosynostosis-specific features. We showcased two
original craniosynostosis-specific applications of our model. Second, we
presented craniosynostosis classification pipeline using the parameter vector
of an \ac{ssm}. We achieved state-of-the-art results comparable to both
\ac{ct} data and photogrammetric scans and to the best of our knowledge, we
tested it on the largest craniosynostosis-specific dataset to date.

\section*{Declaration of Competing Interest}\label{sec:compInt}

The authors declare that they have no known competing financial interests or
personal relationships that could have appeared to influence the work reported
in this paper.

\section*{CRediT author statement} \textbf{Matthias Schaufelberger:}
Conceptualization, Methodology, Software, Validation, Formal analysis,
Investigation, Data curation, Writing - original draft, Visualization.
\textbf{Reinald Peter Kühle:} Conceptualization, Methodology, Investigation,
Resources, Data curation, Writing - review \& editing. \textbf{Andreas
Wachter:} Conceptualization, Methodology, Writing - Review \& Editing.
\textbf{Frederic Weichel:} Conceptualization, Investigation, Resources, Data
curation. \textbf{Niclas Hagen:} Conceptualization, Resources, Data curation.
\textbf{Friedemann Ringwald:} Resources, Data curation. \textbf{Urs
Eisenmann:} Conceptualization, Methodology, Writing - review \& editing,
Supervision, Project administration, Funding acquisition. \textbf{Jürgen
Hoffmann:} Resources, Supervision, Project administration, Funding
acquisition. \textbf{Michael Engel:} Resources, Supervision, Project
administration. \textbf{Christian Freudlsperger:} Conceptualization,
Methodology, Investigation, Resources, Writing - review \& editing,
Supervision, Project administration, Funding acquisiiton. \textbf{Werner
Nahm:} Conceptualization, Methodology, Investigation, Writing - review \&
editing, Supervision, Project administration, Funding acquisition.

\section*{Acknowledgments}

\begin{sloppypar}
This work received funding by the HEiKA research project grant \mbox{HEiKA\_19--17}.
The authors acknowledge the support by the state of Baden-Württemberg through
bwHPC. The authors would like to thank the reviewers for their valuable
comments to improve their work.
\end{sloppypar}

\appendix

\section{Description of \acf{osnicp} methods}\label{ap:osnicpMethods}

To be consistent with the notation in the original paper~\cite{amberg2007}, we 
change notation. The $n_\mathrm{p}$ template points are expressed as 
$\mathbf{V} \in \mathbb{R}^{n_\mathrm{p} \times 3}$. 

The unknown affine transformations are defined as $\mathbf{X} \in \mathbb{R}
^{4n_\mathrm{p} \times 3}$. The full cost function can be expressed as
$E(\mathbf{X}) = \alpha E_\textrm{s}(\mathbf{X}) + E_\textrm{d}(\mathbf{X}) +
\beta E_\textrm{l}(\mathbf{X})$. The stiffness term $E_\mathrm{s}
(\mathbf{X})$ can be described as the Kronecker product $\otimes$ of the mesh
topology matrix $\mathbf{M} \in \mathbb{R}^{n_\textrm{e} \times n_\textrm{p}}$
with $n_\textrm{e}$ denoting the number of edges and $n_\textrm{p}$ the number
of points. The weight matrix $\mathbf{G} \in \mathbb{R}^{4 \times 4} =
\diag(1,1,1,\gamma)$ between rotational and skew parts against translational
parts~\cite{amberg2007}:

\begin{equation}
    E_\mathrm{s} (\mathbf{X}) = \norm{ ( \mathbf{M} \otimes \mathbf{G}) \mathbf{X} } ^2_F.
    \label{eq:stiffness}
\end{equation}

$\mathbf{M}$ describes the connections between neighboring vertices (we use 
the node-arc incidence matrix~\cite{jeter1986} in which for each edge $r$ 
we set $\mathbf{M}(r,i) = -1$ and $\mathbf{M}(r,j) = 1$). The distance term 
$E_\mathrm{d} (\mathbf{X}$) describes how close the displaced template vertices 
are to the target vertices and can be written as: 

\begin{equation}
    E_\mathrm{d} (\mathbf{X}) = \norm{ \mathbf{W} (\mathbf{D} \mathbf{X} - \mathbf{U}) } ^2_F.
    \label{eq:distance}
\end{equation}

$\mathbf{W} \in \mathbb{R} ^ {n_\mathrm{p} \times n_\mathrm{p}}$ is a diagonal
weighting matrix which allows assigning different weights to each
transformation. The sparse displacement matrix $\mathbf{D} \in \mathbb{R} ^
{n_\mathrm{p} \times 4n_\mathrm{p}}$ is a diagnoal matrix with the homogeneous
points $v_i= [x_i, y_i, z_i, 1]^\mathtt{T}$ as its diagnoal elements mapping
the homogeneous template points to the respective affine transforms.
$\mathbf{U} \in \mathbb{R} ^ {n_\mathrm{p} \times 3}$ denotes the found
correspondences from the target points. 

Finally, the landmark term $E_\mathrm{l} (\mathbf{X})$ is similar to the 
distance term while only the landmark points are considered: 

\begin{equation}
    E_\mathrm{l} (\mathbf{X}) = \norm{ (\mathbf{D}_\mathrm{L} \mathbf{X} - 
    \mathbf{U}_\mathrm{L}) } ^2_F.
    \label{eq:landmark}
\end{equation}

The complete cost function for \ac{nicpa} can be written as: 

\begin{equation}
    E(\mathbf{X}) = 
    \left\lVert
    \begin{bmatrix}
    \alpha \mathbf{M} \otimes \mathbf{G} \\
    \mathbf{W} \mathbf{D} \\
    \beta \mathbf{D}_\mathrm{L}
    \end{bmatrix}
    \mathbf{X} - 
    \begin{bmatrix}
    \mathbf{0} \\
    \mathbf{W} \mathbf{U} \\
    \mathbf{U}_\mathrm{L}
    \end{bmatrix}
    \right\rVert ^2_F
    \label{eq:fullcostfunction}
\end{equation}

For the translation-only variant \ac{nicpt}, the unknown transformations are
defined as translations $\mathbf{X} \in \mathbb{R} ^{n_\mathrm{p} \times 3}$.
The cost function is changed accordingly:

\begin{equation}
    E(\mathbf{X}) = 
    \left\lVert
    \begin{bmatrix}
    \alpha \mathbf{M} \\
    \mathbf{W} \mathbf{I}_{n_\mathrm{p}}
    \end{bmatrix}
    \mathbf{X} - 
    \begin{bmatrix}
    \mathbf{0} \\
    \mathbf{W} (\mathbf{U} - \mathbf{V})
    \end{bmatrix}
    \right\rVert ^2_F
    \label{eq:fullcostfunctionTranslation}
\end{equation}

\section{Description of \acl{lbrp} methods}\label{ap:lbrpMethods}

\Acf{lbrp}~\cite{dai2017a,dai2020} relies on mutual correspondences between 
template and target and uses the \ac{lb} operator $\mathbf{L}_\mathrm{0} \in 
\mathbb{R} ^ {n_\mathrm{p} \times n_\mathrm{p}}$ computed on the original 
template as a regularization, controlled by the stiffness parameter $\lambda$.  
A higher $\lambda$ puts more weight to the \ac{lb} term of the equation, 
leading to a mesh which retains its original shape. For a low $\lambda$, the 
original template shape is disregarded and is mapped closer to the target 
mesh, which might lead to irregularities in the projection. This template 
projection step can be described using~\cite{dai2020,dai2017a}:

\begin{equation}
\begin{bmatrix}
\lambda \mathbf{L}_\mathrm{0} \\
\mathbf{S}_\mathrm{X}
\end{bmatrix}
\mathbf{X} =
\begin{bmatrix}
\lambda \mathbf{L}_\mathrm{0}\mathbf{X}_\mathrm{0}\\
\mathbf{S}_\mathrm{Y} \mathbf{Y}
\end{bmatrix},
\label{eq:adaptive}
\end{equation}

The two Boolean selection matrices $\mathbf{S}_{\mathrm{X}} \in 
\interval{0}{1}^{k \times n_\mathrm{n}}$ and $\mathbf{S}_{\mathrm{Y}} \in 
\interval{0}{1}^{k \times n_\mathrm{t}}$ select the $k$ correspondences on the 
template and target. $n_\mathrm{p}$ denotes the number of template points, 
$n_\mathrm{t}$ the number of target points.

For the \ac{2slbrp}, we essentially perform this template projection twice, 
first to adapt the template to the target and then refining it with decreased 
stiffness to let it deform more strongly to the target. This is a similar
approach to the template adaption described in~\cite{dai2020}.

For the \ac{icpdlbrp}~\cite{dai2020}, we first employ the adaptive template 
projection with a high $\lambda$, then alternate between the rigid \ac{cpd} 
and nonrigid \ac{cpd}~\cite{myronenko2010} and conclude with the adaptive 
template projection with a small $\lambda$. For our dataset, this increased 
robustness compared to using the affine \ac{cpd} variant as proposed 
in~\cite{dai2020}. For further reading on the \ac{icpd} and the template 
adaption of the \ac{lbrp}, the reader is referred to~\cite{dai2020}.

\section{Evaluation of the morphing methods}\label{ap:morphingComparison}

\begin{figure}[htbp!]
\begin{subfigure}[t]{0.32\textwidth}
\centering
\includegraphics[width=\textwidth]{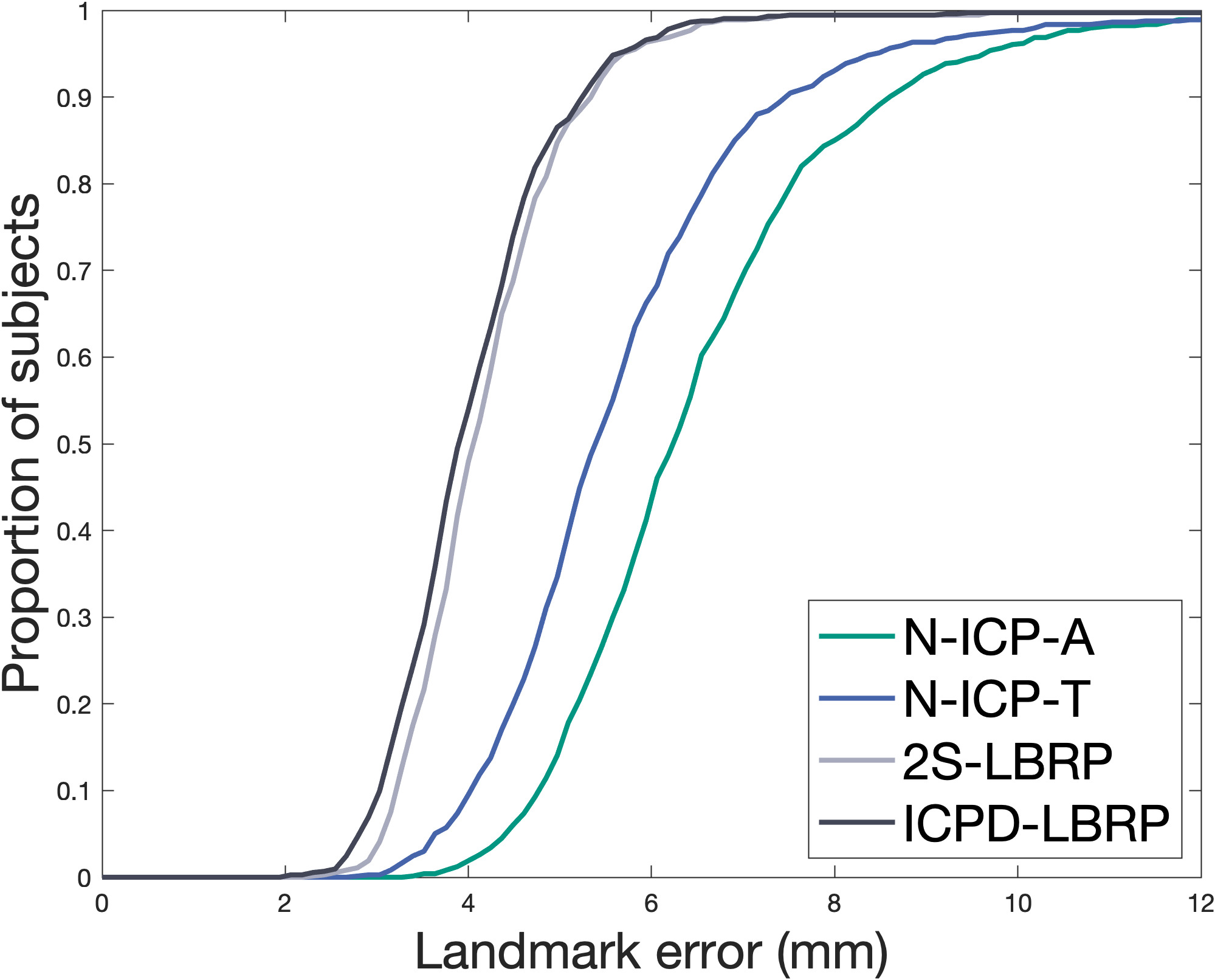}
\end{subfigure}
\begin{subfigure}[t]{0.32\textwidth}
\centering
\includegraphics[width=\textwidth]{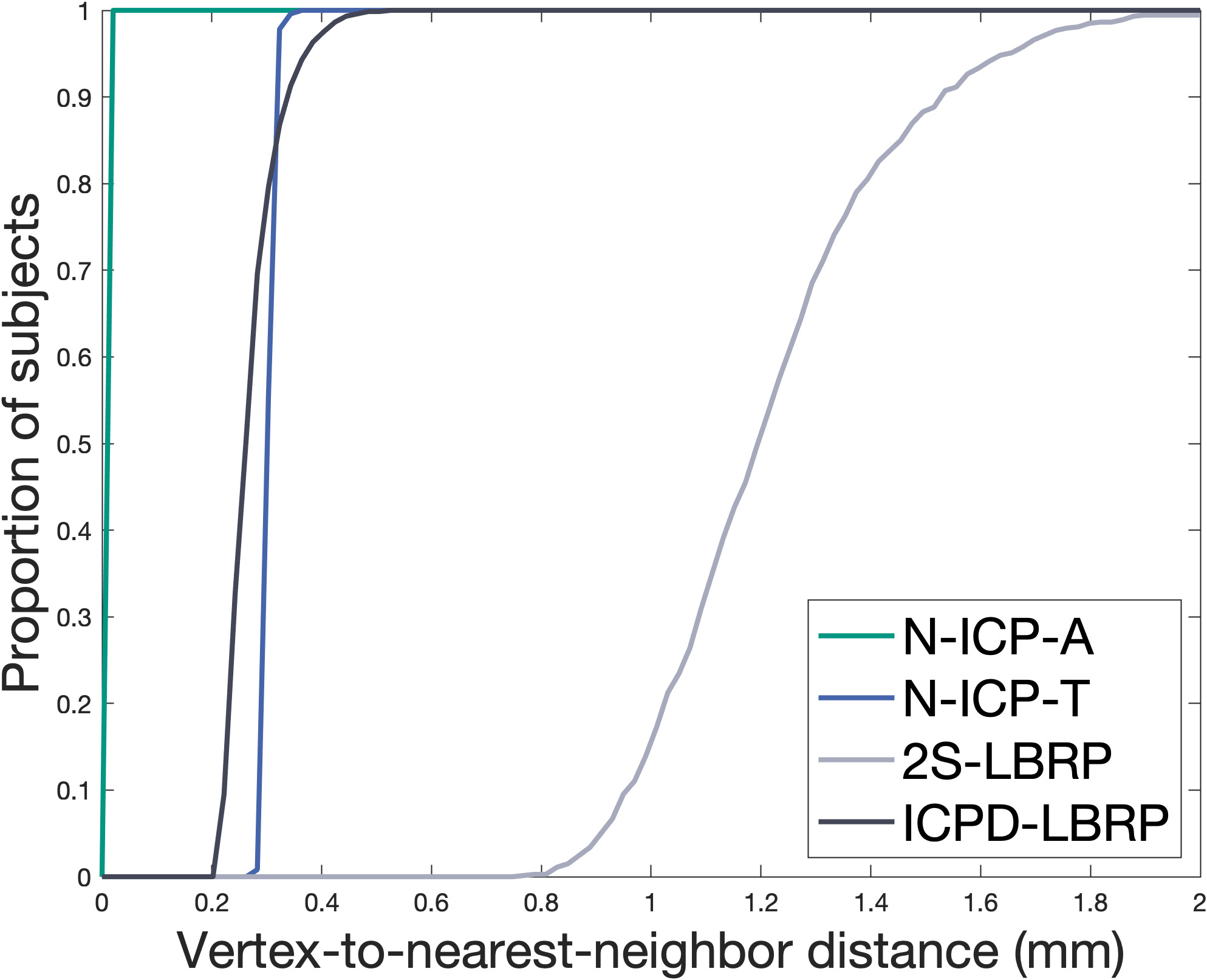}
\end{subfigure}
\begin{subfigure}[t]{0.32\textwidth}
\centering
\includegraphics[width=\textwidth]{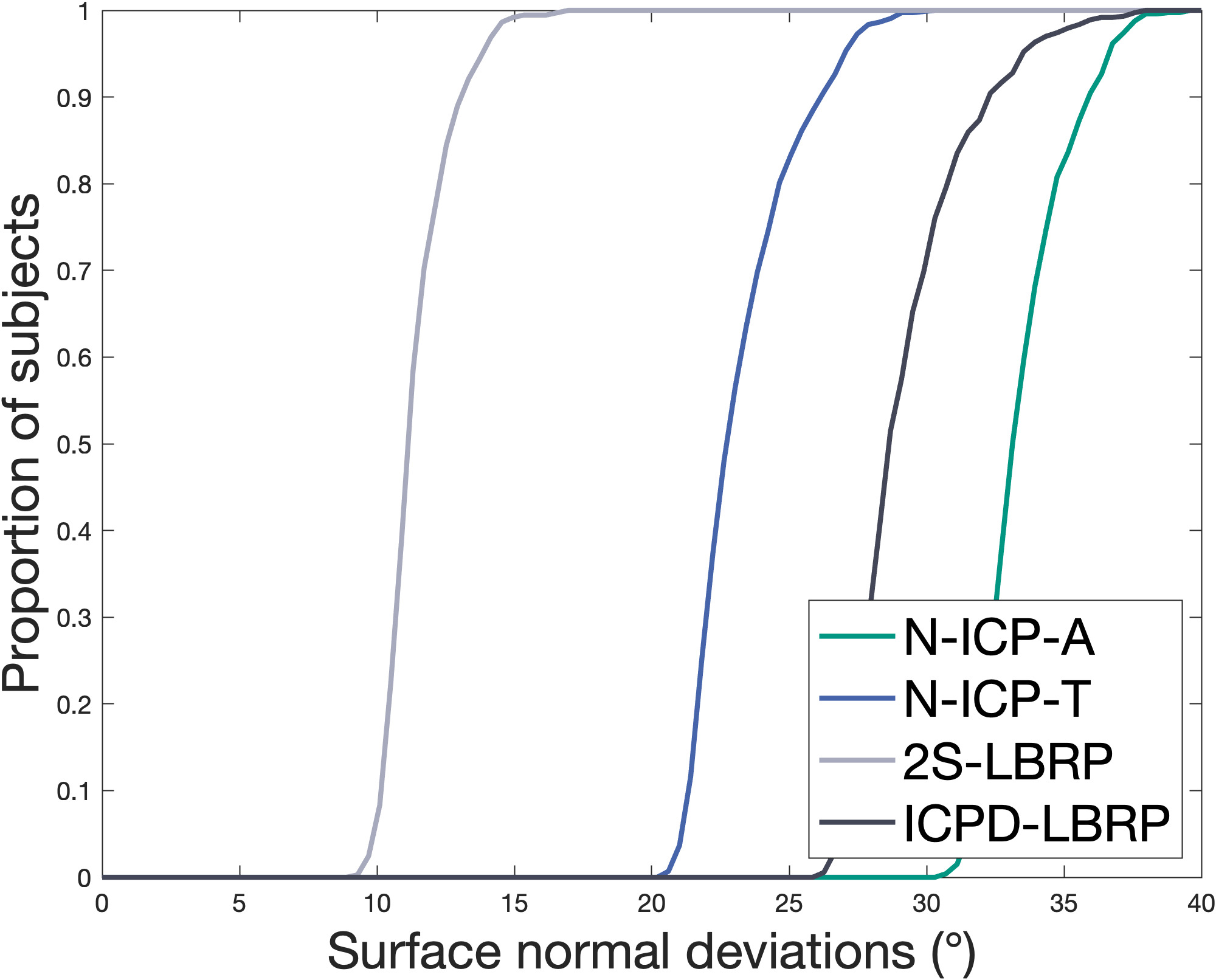}
\end{subfigure}
\caption{Proportion of subjects with mean landmark error,
vertex-to-nearest-neighbor-distances, and surface normal. We compared
\acl{nicpa}, \acl{nicpt}, \acf{2slbrp}, and \acf{icpdlbrp}. The graph shows
the proportion of subjects less than the abscissa value. Higher is better.}\label{fig:metricsFull}
\end{figure}

We present mean and standard deviations for each error metric
in~\Cref{tab:morphingmean}. Cumulative errors for each metric show the
distribution of each error and are displayed in~\Cref{fig:metricsFull}.
\Ac{lbrp} methods showed smaller landmark errors and larger surface normal
deviations compared to \ac{osnicp}. \Ac{nicpa} had the lowest
vertex-to-nearest-neighbor distance errors. Surface normal deviations were 
for all methods larger than $17\degree$.

\begin{table}[htbp!]
\caption{Mean error and standard deviation for each morphing method. Boldface
shows smallest error for each metric.}\label{tab:morphingmean}
\centering
\small
\begin{tabularx}{0.99\textwidth}{Xzzz}
\toprule
Morphing method & Mean landmark error (mm) & Mean vertex-to-nearest-neighbor distance (mm) & Mean surface normals deviations (degree) \\
\midrule
\midrule
\Acf{nicpa}    & $6.533 \pm 1.796$          & $\mathbf{0.007 \pm 0.003}$ & $33.488 \pm 1.578$          \\ \midrule
\Acf{nicpt}    & $5.699 \pm 1.789$          & $0.302 \pm 0.01$           & $23.242 \pm 1.849$          \\ \midrule
\Acf{2slbrp}   & $4.185 \pm 1.205$          & $0.785 \pm 0.17$           & $\mathbf{20.392 \pm 1.466}$ \\ \midrule
\Acf{icpdlbrp} & $\mathbf{4.071 \pm 1.163}$ & $0.272 \pm 0.049$          & $29.255 \pm 2.12$           \\
\bottomrule
\end{tabularx}
\end{table}

\newcommand{\compheightAll}{0.29}
\begin{figure}[ht]
    \centering
    \includegraphics[height=\compheightAll\textwidth]{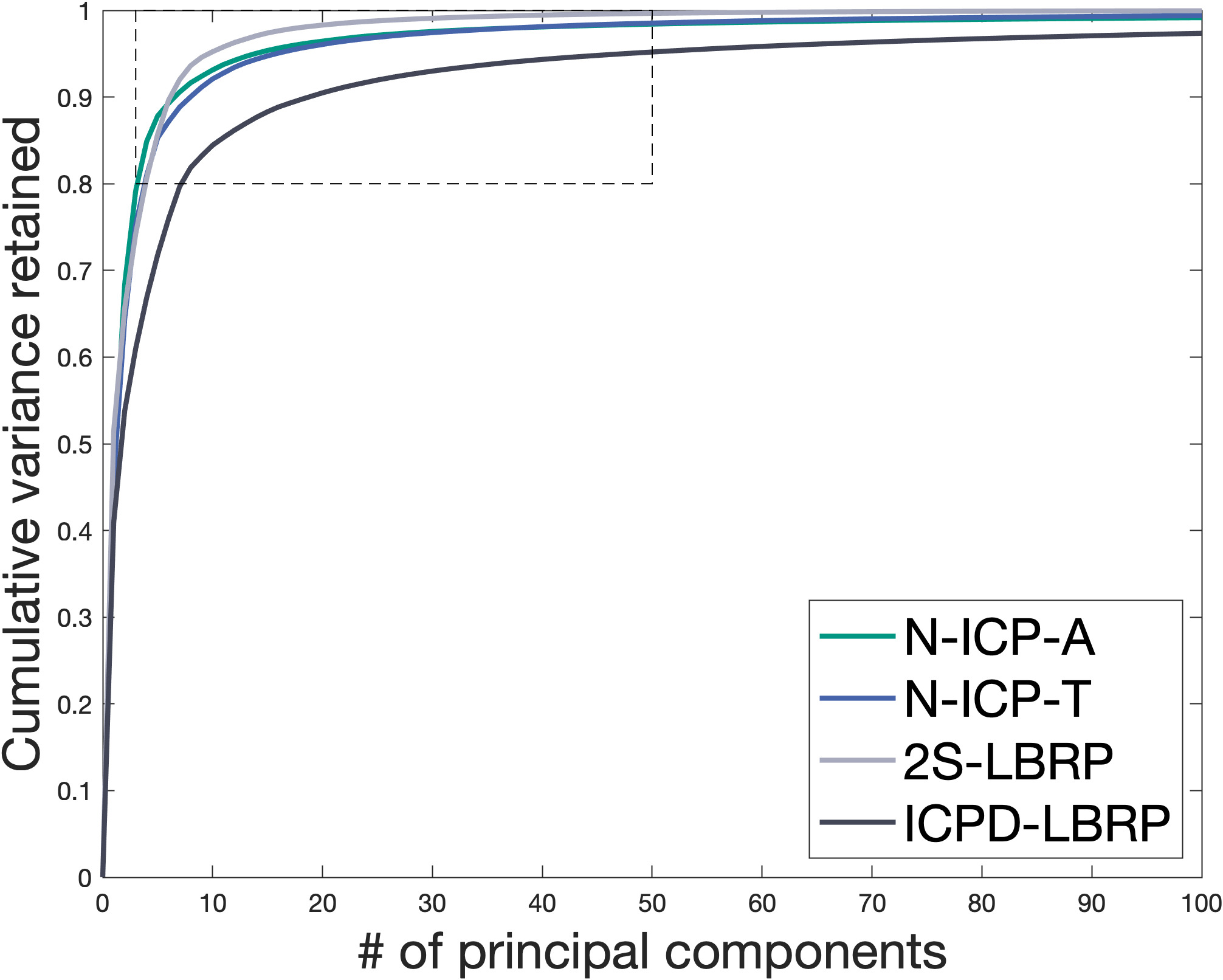}
    \includegraphics[height=\compheightAll\textwidth]{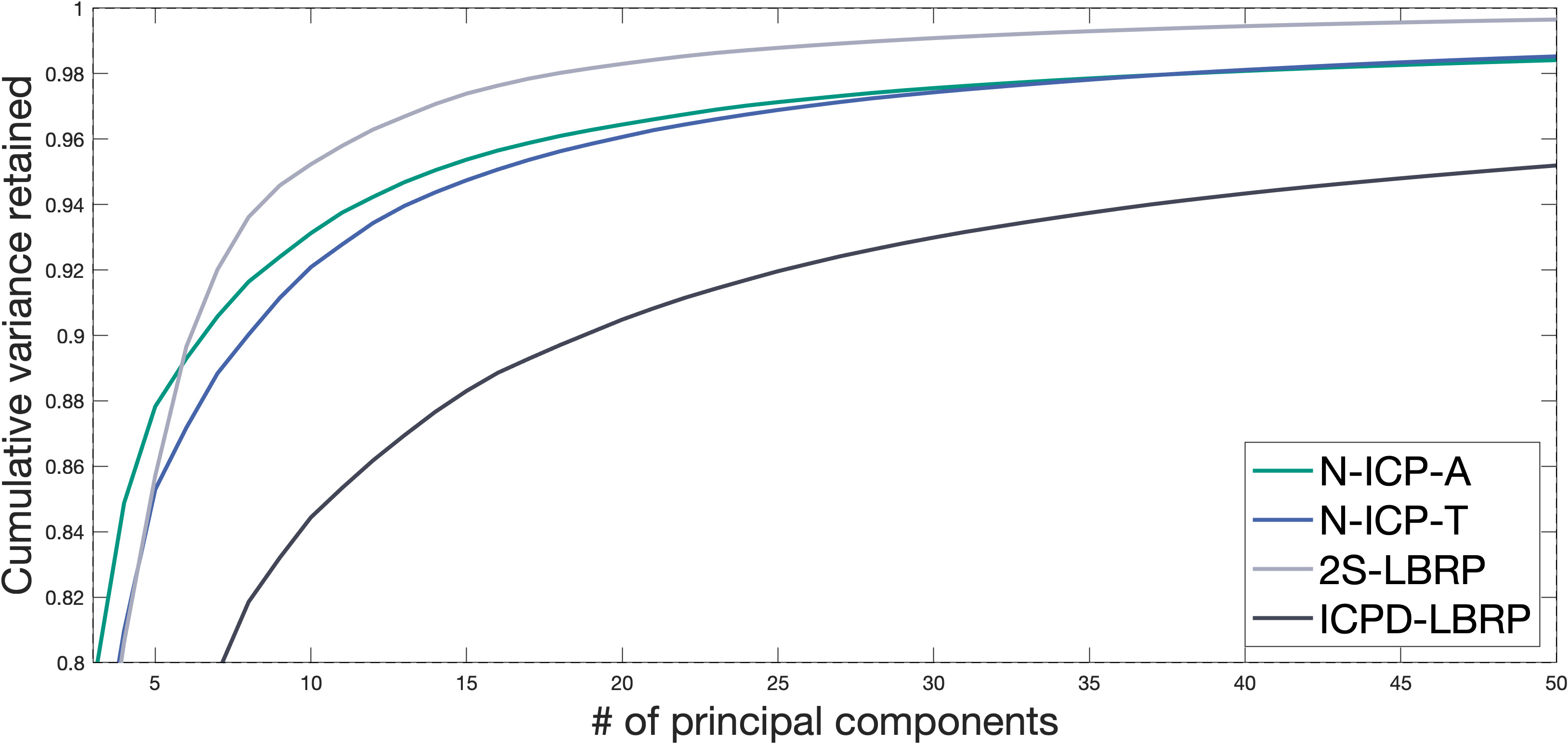}
    \caption{Compactness as a function of the number of principal components
    of the full shape model. We compared \acf{nicpa}, \acf{nicpt}, \acf{2slbrp},
    and \acf{icpdlbrp}. Left: full compactness, right: zoom-in. A higher value
    is better.}
    \label{fig:compactness}
\end{figure}

\begin{figure}[htbp!]
\begin{subfigure}[c]{0.49\textwidth}
\centering
\includegraphics[width=\textwidth]{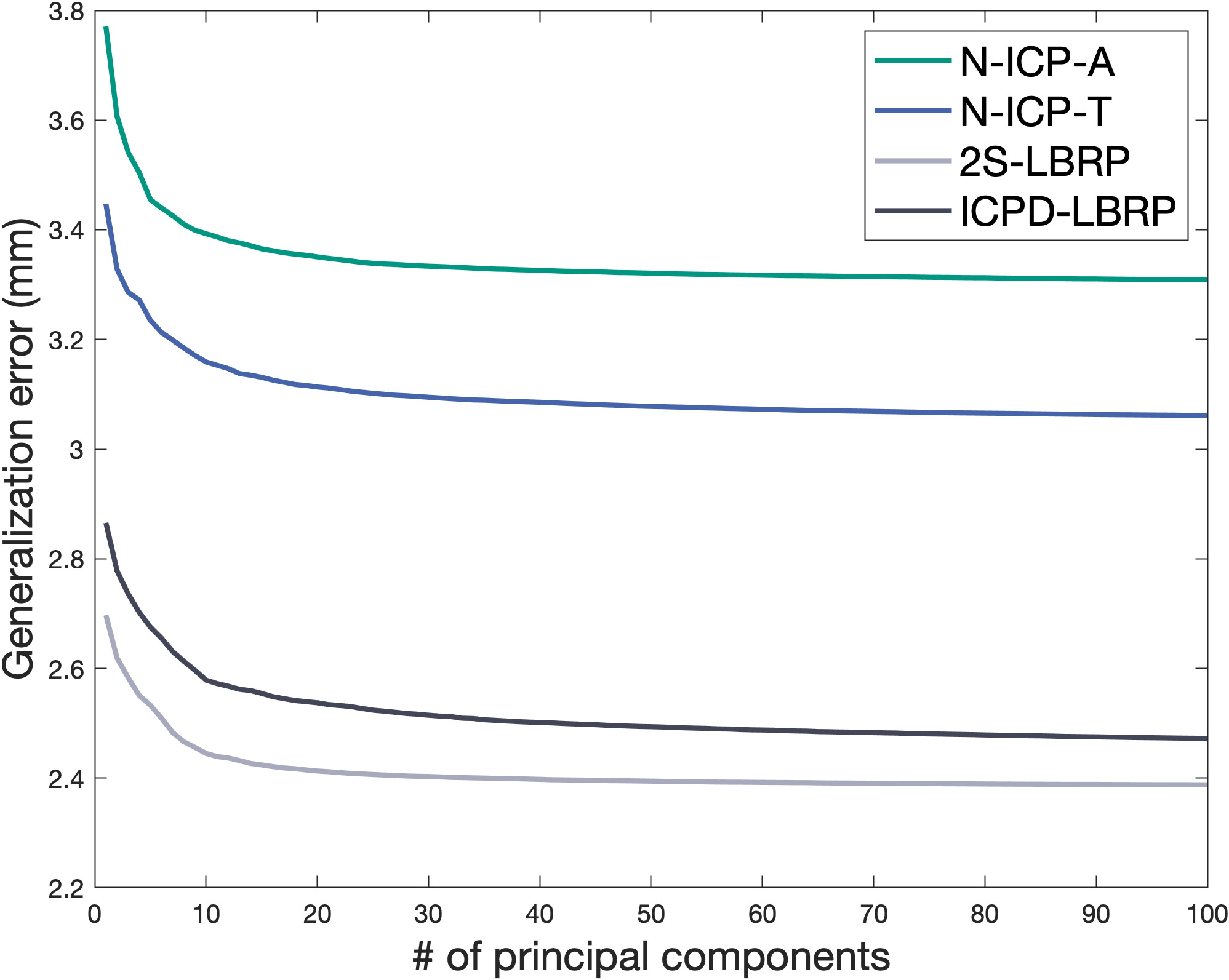}
\end{subfigure}
\begin{subfigure}[c]{0.49\textwidth}
\centering
\includegraphics[width=\textwidth]{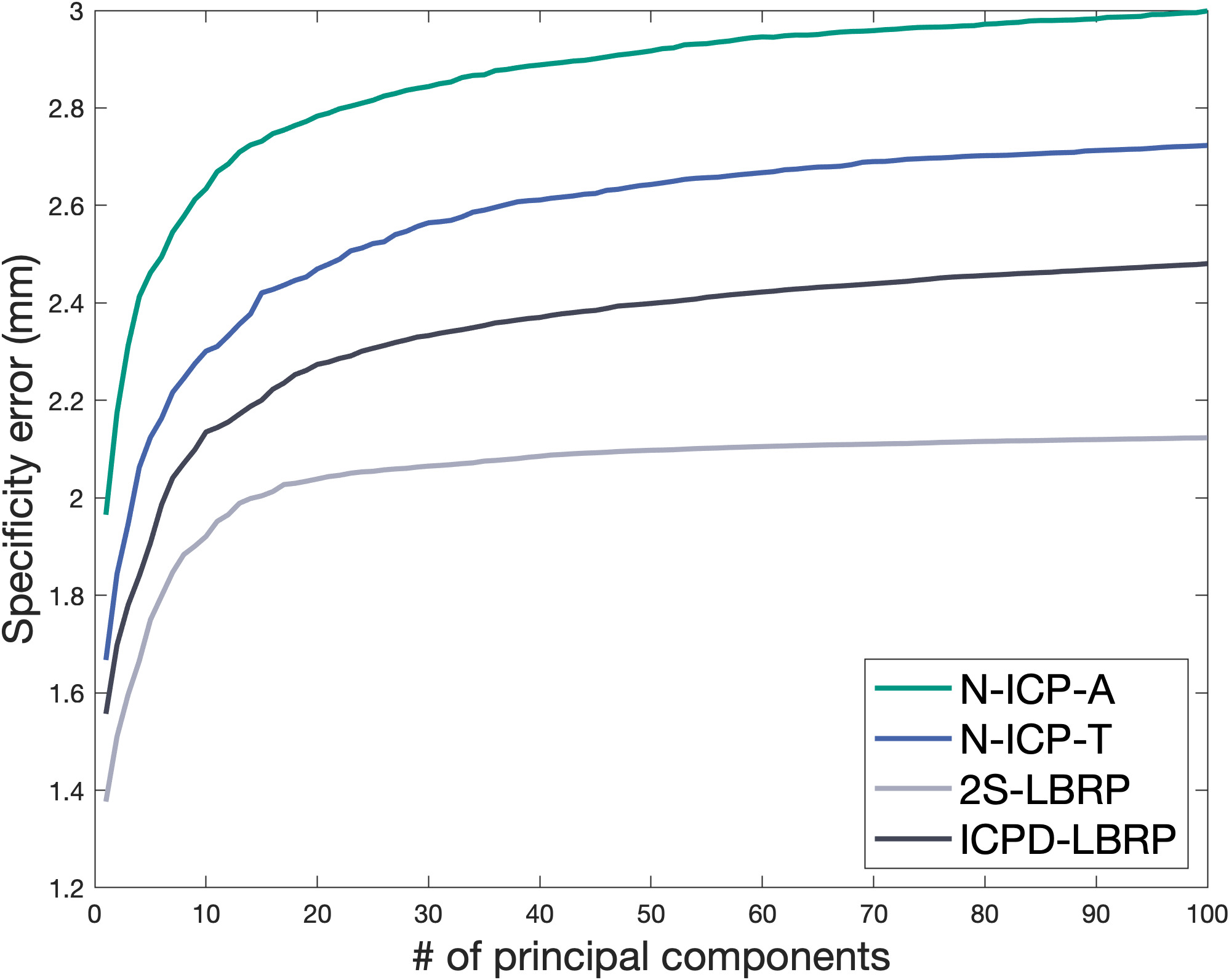}
\end{subfigure}
\caption{Generalization and specificity errors as functions of the number of
principal components of the full shape model. We compared \acf{nicpa},
\acf{nicpt}, \acf{2slbrp}, and \acf{icpdlbrp}. Left: generalization error,
right: specificity error. For both metrics, a lower error is better.}\label{fig:genspe}
\end{figure}

\Cref{fig:compactness} shows the compactness of the \ac{ssm}. The most compact 
models were produced by \ac{nicpa} and \ac{nicpt}, while generalization error 
and specificity error were larger~(\Cref{fig:genspe}).

\section{Comparison of classification results for all four morphing 
methods}\label{ap:morphingClassification}

In \Cref{tab:classifiersMorphing} we show the classification results per
morphing method and classification approach. \Ac{lda} consistently yielded the
highest accuracy regardless of the morphing method. Using \ac{lda}, all
morphing methods obtained accuracies of 97.0\,\% or higher. Note that 
\ac{nicpt} model scores the highest accuracy although it did not perform best 
in any of the evaluation criteria.

\begin{table}[ht]
\caption{Highest accuracy for each classifier and morphing methods.
Optimal number of principal components is given in brackets. Underlined
accuracy indicates optimal classifier per method and boldface optimal
classifier overall.}\label{tab:classifiersMorphing}
\centering
\small
\begin{tabularx}{\textwidth}{Xttttt}
\toprule
Morphing \parbox{3cm}{method} & \ac{lda} & \ac{svm} & \ac{nb} & \ac{knn} & \ac{bdt} \\
\midrule
\midrule
\Acl{nicpa}    & \underline{0.978}  (44) & 0.962  (46) & 0.946  (23) & 0.916  (23) & 0.826  (22) \\ \midrule
\Acl{nicpt}    & \textbf{0.981}  (54)    & 0.959  (23) & 0.946  (28) & 0.910  (9)  & 0.842  (10) \\ \midrule
\Acl{2slbrp}   & \underline{0.970} (50)  & \underline{0.970}  (31) & 0.959  (23) & 0.946  (10) & 0.861  (8) \\ \midrule
\Acl{icpdlbrp} & \underline{0.975}  (91) & 0.962  (13) & 0.951  (23) & 0.951 (10) & 0.847  (8) \\
\bottomrule
\end{tabularx}
\end{table}

\section{Landmarks}\label{ap:landmarks}

\Cref{tab:landmarks} shows the landmarks annotated by the medical staff.
\begin{table}[htbp!]
\caption{Landmarks on 3D surface scans provided by the medical staff. We use the cephalometric landmark notation of~\cite{swennen2006}.}\label{tab:landmarks}
\small
\centering
\begin{tabularx}{0.8\textwidth}{Xh}
\toprule
Landmark & Abbreviation\\
\midrule
\midrule
Tragion (left and right) & (\(t_\mathrm{l}\)) and (\(t_\mathrm{r}\))\\
Sellion & (\(se\))\\
Exocanthion (left and right) & (\(ex_\mathrm{l}\)) and (\(ex_\mathrm{r}\))\\
Subnasale & (\(sn\))\\
Labiale Superius & (\(ls\))\\
Otobasion superius (left and right) & (\(obs_\mathrm{l}\)) and (\(obs_\mathrm{r}\))\\
Soft tissue gnathion & (\(gn\))\\
\bottomrule
\end{tabularx}
\end{table}

\section{Hyperparameters for template morphing}\label{ap:hyperparameters}

\Cref{tab:morphinghyper} lists the hyperparameters used in each method.
\begin{table}[htbp!]
\caption{Hyperparameters used for the template morphing approaches.}\label{tab:morphinghyper}
\centering
\small
\begin{tabularx}{\textwidth}{XX}
\toprule
\multicolumn{2}{l}{\Acf{2slbrp} (notation of~\cite{dai2020}) } \\
Stiffness first morph & $\lambda_1$ = 10 \\
Stiffness second morph & $\lambda_2$ = 0.1 \\
\midrule
\midrule
\multicolumn{2}{l}{\Acf{icpdlbrp}} \\
\multicolumn{2}{l}{(notation of~\cite{dai2020}) } \\
Stiffness first morph & $\lambda_1$ = 10 \\
\Acf{icpd}-Loop & For each iteration, perform first \textit{cpdRigid}, then
\textit{cpdNonrigid} \\
\quad \textit{cpdNonrigid} smoothing weight: & 3\\
\quad \textit{cpdNonrigid} tolerance & $1 \cdot 10^{-5}$\\
\quad Exit condition & fewer than 1\,\% of nearest neighbors between
iterations change\\
Stiffness second morph & $\lambda_2$ = 0.1 with Laplace matrix resulting from
first morph\\
\midrule
\midrule
\multicolumn{2}{l}{\Acf{nicpa} and} \\
\multicolumn{2}{l}{\acf{nicpt} (notation of~\cite{amberg2007}) } \\
Iterations & $n$ = 80 \\
\quad Stiffness parameter $\alpha$ in iteration $n$ & $\alpha_n = 10^8 \cdot 0.8^{n}$ \\
\quad Landmark weight in iteration $n$ $\beta$ & if $n<51$ $\beta_{n} = 1$, else $\beta_{n} = 0$ \\
\quad Exit condition $\epsilon$ for each fixed stiffness $\alpha$ & $\epsilon < 100$ \\
Valid normals for correspondence establishment $\varphi$ & $\varphi < 45 \degree$ \\
Rotation weight $\gamma$ & $\gamma = 1$ \\
\bottomrule
\end{tabularx}
\end{table}

\clearpage

\bibliography{references_full}

\end{document}